\documentclass[twocolumn]{aastex62}

\usepackage{float}
\usepackage{ulem}
\usepackage{mathrsfs}
\usepackage{amsmath}
\begin{document}

\newcommand{\gaia}{{\em Gaia}}
\newcommand{\kms}{{\rm{km\ s^{-1}}}}
\newcommand{\msun}{{M_\odot}}
\newcommand{\yr}{{{\rm{yr}}}}
\newcommand{\rapid}{{{\tt rapid}}}
\newcommand{\delayed}{{{\tt delayed}}}
\newcommand{\porb}{{{P_{\rm{orb}}}}}
\newcommand{\metal}{{{\rm Z}}}
\newcommand{\mlc}{{{M_{\rm{LC}}}}}
\newcommand{\llc}{{{L_{\rm{LC}}}}}
\newcommand{\myr}{{{\rm Myr}}}
\newcommand{\mbh}{{{M_{\rm{BH}}}}}
\newcommand{\mtot}{{M_{\rm{tot}}}}
\newcommand{\ecc}{{{Ecc}}}
\newcommand{\kpc}{{{\rm{kpc}}}}
\newcommand{\pc}{{{\rm{pc}}}}
\newcommand{\BH}{{\rm{BH}}}
\newcommand{\LC}{{\rm{LC}}}
\newcommand{\macc}{{M_{\rm{acc}}}}
\newcommand{\racc}{{R_{\rm{acc}}}}
\newcommand{\ergspers}{{{\rm{ergs\ s^{-1}}}}}
\newcommand{\cosmic}{{{\texttt{COSMIC}}}}
\newcommand{\dustmaps}{{\texttt{mwdust}}}
\newcommand{\bse}{{{\texttt{BSE}}}}

\title{{\em Gaia} May Detect Hundreds of Well-characterised Stellar Black Holes}

\author[0000-0001-9685-3777]{Chirag Chawla}
\affiliation{Tata Institute of Fundamental Research, Department of Astronomy and Astrophysics, Homi Bhabha Road, Navy Nagar, Colaba, Mumbai, 400005, India}
\email{chirag.chawla@tifr.res.in}

\author[0000-0002-3680-2684]{Sourav Chatterjee}
\affiliation{Tata Institute of Fundamental Research, Department of Astronomy and Astrophysics, Homi Bhabha Road, Navy Nagar, Colaba, Mumbai, 400005, India}
\email{souravchatterjee.tifr@gmail.com}

\author[0000-0001-5228-6598]{Katelyn Breivik}
\affiliation{Center for Computational Astrophysics, Flatiron Institute, 162 Fifth Ave, New York, NY, 10010, USA}

\author[0000-0001-7825-2062]{Chaithanya Krishna Moorthy}
\affiliation{Department of Physics, Indian Institute of Technology Madras, Chennai, 600036, India}

\author[0000-0001-5261-3923]{Jeff J. Andrews}
\affiliation{Center for Interdisciplinary Exploration and Research in Astrophysics (CIERA), 
1800 Sherman Ave., 
Evanston, IL, 60201, USA}

\author[0000-0003-3939-3297]{Robyn E. Sanderson}
\affiliation{Department of Physics and Astronomy, University of Pennsylvania, 209 South 33rd Street, Philadelphia, PA 19104, USA}
\affiliation{Center for Computational Astrophysics, Flatiron Institute, 162 Fifth Ave, New York, NY, 10010, USA}

\begin{abstract}
Detection of black holes (BHs) with detached luminous companions (LCs) can be instrumental in connecting the BH properties with their progenitors’ since the latter can be inferred from the observable properties of the LC. Past studies showed the promise of \gaia\ astrometry in detecting BH-LC binaries. We build upon these studies by: 1) initialising the zero-age binary properties based on realistic, metallicity-dependent star-formation history in the Milky Way (MW), 2) evolving these binaries to current epoch to generate realistic MW populations of BH-LC binaries, 3) distributing these binaries in the MW preserving the complex age-metallicity-Galactic position correlations, 4) accounting for extinction and reddening using three-dimensional dust maps, 5) examining the extended \gaia\ mission’s ability to resolve BH-LC binaries. We restrict ourselves to detached BH-LC binaries with orbital period $\porb\leq10\,\yr$ such that \gaia\ can observe at least one full orbit. We find: 1) the extended \gaia\ mission can astrometrically resolve $\sim30$--$300$ detached BH-LC binaries depending on our assumptions of supernova physics and astrometric detection threshold; 2) \gaia's astrometry alone can indicate BH candidates for $\sim10$--$100$ BH-LC binaries by constraining the dark {\em primary} mass $\geq3\,\msun$; 3) distributions of observables including orbital periods, eccentricities, and component masses are sensitive to the adopted binary evolution model, hence can directly inform binary evolution models. Finally, we comment on the potential to further characterise these BH binaries through radial velocity measurements and observation of X-ray counterparts.
\end{abstract}
\section{Introduction}
\label{S:intro}
The recent discoveries of merging binary black holes (BHs) by the LIGO-Virgo and Kagra observatories have reignited the interest in understanding the astrophysical origins of BH binaries in short-period orbits \citep[e.g.,][]{Abbott2016a, Abbott2016b,GWTC1,GWTC2,GWTC2-pop}. A major hurdle in understanding the astrophysical implications of these detections, as well as modeling realistic populations of binary BHs can be attributed to uncertainties in how massive stars evolve and form compact objects \citep[e.g.,][]{Woosley2002,Fryer2012,Sukhbold2016,Woosley2017,Sukhbold2018,Pejcha2020}. 

It is expected that there are $\sim 10^7-10^9$ stellar-mass BHs in the Milky Way \citep[e.g.,][]{Brown1994,Timmes1996,Samland1998,Olejak2020}. However, discovering them using traditional methods such as X-ray or radio observations is notoriously difficult resulting in only $\sim60$ detections to date \citep[e.g.,][]{Remillard2006}. Only a small fraction of BH binaries are expected to be actively accreting at a detectable rate at any given time due to the stringent requirements on the orbital and stellar properties of accreting systems and their typically low duty cycles \citep{Fragos2009,Fabbiano2012,Gallo2014,Corral-Santana2016,Tetarenko2016}. Furthermore, the observed population of BHs in low mass X-ray binaries may be biased toward lower masses due to observational selection effects \citep{Jonker2021}.

The population of BHs detected via gravitational waves (GWs) emitted from the inspiral and merger of binary BHs also have strong selection biases favoring distant and high-mass objects \citep[e.g.,][]{Fishbach2017}. Thus, the historically standard methods for BH detections likely miss the bulk of the population of BH binaries in the Milky Way. Even when discovered, GW detections do not directly constrain age and metallicity of the progenitors of the merging BHs. Instead, such constraints come indirectly from population modeling of various astrophysical formation channels \citep[e.g.,][]{Chatterjee2017-metal,GWTC2,GWTC2-pop,Bavera2021,Chruslinska2021}. On the other hand, detailed modeling of complex accretion physics is needed to characterise the BH and the companion properties in X-ray binaries since the properties of the accretion disks are the observables in this case \citep[e.g.,][]{FKR-book, Davis2005, Kreidberg2012}. 

The majority of BH binaries are expected to have stellar companions in orbits too wide for mass transfer via Roche-lobe overflow (RLOF) \citep[e.g.,][]{Breivik_2017}. Detection of a detached BH binary with a luminous companion (LC) is interesting since such a source can effectively remove several limitations of the aforementioned methods. In a detached BH-LC binary, the properties, such as the age and metallicity of the BH progenitor can be assumed if the same properties can be measured for the LC. For example, knowledge of the brightness and distance of the LC can allow strong constraints for the luminosity and metallicity using well-understood  stellar evolution models. If the BH-LC binary is primordial, as expected for most BH binaries in the Milky Way field, the metallicity and age of the BH progenitor is very likely the same as its companion. Furthermore, especially if the LC is a main-sequence (MS) star, the knowledge of luminosity and color can provide strong constraints on the mass of the star and as a result, the mass of the dark object can also be constrained \citep[e.g.,][]{Andrews_2019, Shikauchi2020}. Interestingly, a fraction of detached BH-LC binaries in the field may also have been created dynamically inside star clusters which then get ejected from the host cluster \citep[e.g.,][]{Chatterjee2017,Kremer2018}. Since star clusters consist of an effectively co-eval aggregate of stars, the same assumptions should be valid for dynamically produced systems as well  to a large degree.\footnote{Some degree of uncertainty may stem from the usually small spread in metallicity and age created via multiple populations in star clusters \citep[e.g.,][]{Milone2020} depending on how much stars from different populations mix and take part in strong dynamical encounters.}    

Detached BH-LC binaries are almost impossible to detect via traditional methods such as radio, GW, or X-ray emissions, with the only exception being wind-fed X-ray binaries. Nevertheless, several BH candidates in detached binary systems with a LC have already been detected by multi-epoch spectroscopic and photometric campaigns via measurement of the orbital motion of the LC \citep[e.g.,][]{Giesers2018,Giesers2019,Thompson2019,Jayasinghe2021}. Systematic surveys for radial velocities of all nearby stars by APOGEE or SDSS-V may provide more discoveries in future \citep[e.g.,][]{Zasowski2017, Kollmeier2019}.

Recently, several groups have proposed that large numbers of detached BH-LC binaries may be detected by resolving the orbital motion of LCs around dark objects using astrometry by \gaia\ \citep{Gould2002,Barstow2014,Mashian2017,Breivik_2017,Yalinewich2018,Yamaguchi2018,Breivik2019,Wiktorowicz2020}. Since \gaia\ will provide position and parallax measurements to $\sim\mu\rm{as}$ precision, it will be relatively straightforward to estimate the LC's luminosity and temperature from magnitude and color without the need for additional followup observations. As a result, the age, metallicity, and mass of the LCs (especially if the LC is a main-sequence star) can be constrained using well-understood stellar modeling \citep[e.g.,][]{Anders2019,Howes2019}. 
In addition, a population of BHs detected this way, arguably will constitute the least biased detected to date, since all selection effects in this case depend primarily on the properties of the LC and not directly on the properties of the BH.  

All of the aforementioned studies show that the astrometric motion of a large number of LCs in orbit around unseen dark objects should be resolved using \gaia. 
These studies also illustrate that while the basic idea is robust, the actual yield and the properties of the detected BH-LC binaries can vary widely depending on the assumptions for stellar evolution and binary interactions \citep[e.g.,][]{Breivik_2017}, and how carefully one considers observability and selection effects. Hence, it is crucial to carefully consider the details related to the synthesis of the population of BH-LC binaries as well as their observation by \gaia. In this study we update our earlier works presented in \citet{Breivik_2017}, \citet{Breivik2019}, and \citet{Andrews_2019} by making several improvements in our binary population synthesis and observational considerations. In contrast to these previous works, we now include realistic stellar distributions in the Milky Way, and location-dependent stellar ages and metallicities based on the Ananke Framework of the Latte Suite of the FIRE-2 Simulations \citep{Wetzel2016,Hopkins2018,Sanderson2020}. Furthermore, we consider observational selection effects more accurately by taking into account the number of planned \gaia\ transits for each system and three dimensional extinction, both of which depend on the Galactic positions of the BH-LCs. 

The paper is organised as follows. In \autoref{S:methods} we detail our numerical setup for BH-LC population synthesis and construction of synthetic Milky Way models. In \autoref{S:detectability} we describe how we determine whether a BH-LC binary would be resolvable by \gaia. Here we take into account astrometric detectability, interstellar extinction and reddening, and possibility of mass constraints of the dark objects using astrometry alone. In \autoref{S:results} we present our key results for the population of  BH-LCs resolvable by \gaia's astrometry. In \autoref{S:beyond} we explore promising avenues for followup studies. 
Finally, we summarize our results and conclude in \autoref{S:discussion}.   

\section{Numerical Setup}
\label{S:methods}
To study the properties of the BH-LC population \gaia\ may observe, we generate representative present day BH-LC binary populations using \texttt{COSMIC} \citep{Breivik2020}. \texttt{COSMIC} is a Python-based binary population synthesis suite that employs modified versions of the single and binary star evolution codes \texttt{SSE/BSE} \citep{Hurley2000,Hurley2002}. For a detailed discussion of the modifications to assumptions for binary interaction physics beyond the standard \texttt{BSE} release, see \cite{Breivik2020, Rodriguez2021}. We detail the process to generate a representative Galactic population of BH-LC binaries in the following subsections. 

\subsection{Initialising the binary population}
\label{S:initialising}
\cosmic\ generates binary populations from Zero Age Main Sequence (ZAMS) by assigning each system with an initial age, metallicity, primary mass, secondary mass, orbital period, and eccentricity. We assign ages and metallicities for each binary based on the distribution of star particles in the final snapshot of the simulated galaxy \textbf{m12i} in the Latte suite of the Feedback In Realistic Environments 2 (FIRE-2)\footnote{http://fire.northwestern.edu} simulation suite \citep{Wetzel2016,Hopkins2018}. We use the \textsf{ananke} framework to assign three-dimensional Galactic positions for each binary by resampling from the smoothed density distribution of the star particles, as described in \autoref{S:galactic-realisation} of \citet{Sanderson2020}. These new star formation history assumptions are a major upgrade to our earlier work \citep{Breivik_2017} which assumed constant star formation rate and discrete metallicity values for the thin and thick disks. For a detailed discussion of the resemblance of the adopted star-formation history to that of the Milky Way see \citet{Sanderson2020}. Since the stellar evolution fits employed in \bse\ and thus \cosmic\ are valid for metallicities in the range between $\log(\metal/\metal_\odot)=-2.3$ and $0.2$, we limit the metallicity of our binaries to fall within this range and assign any metallicities outside of the range to the limiting values. 

We assign ZAMS orbital parameters of the binary population by sampling primary masses, secondary masses, orbital periods, and eccentricities from observationally motivated probability distribution functions. In particular, the primary masses are drawn from the \citet{Kroupa2001} initial stellar mass function (IMF) while the secondary companions are assigned masses following a uniform distribution of mass ratios with a range in mass from $ 0.08\,\msun$ to the primary mass \citep{Mazeh1992, Goldberg1994}. Initial eccentricities and orbital periods are drawn by first sampling eccentricities from a thermal distribution \citep{Heggie1975} then sampling semimajor axes that are uniform in log space up to $10^5\,\rm{R}_{\odot}$ \citep{Han1998}. We reject any samples that produce binaries where one of the component stars fills more than half of its Roche radius. Finally, we assume that the initial binary fraction is $0.5$ which places two of every three stars formed into a binary system. 

\subsection{Binary stellar evolution}
\label{S:BSE}
We use \texttt{COSMIC} to simulate binary evolution from ZAMS through to the present day with ages and metallicities assigned by the star particles in galaxy \textbf{m12i}. The present-day orbital characteristics of the BH-LC population depend strongly on the assumptions for the outcomes of RLOF mass transfer from the BH progenitor as well as natal kicks which can be imparted to the BH during its formation. We parameterize the stability of mass transfer using critical mass ratios defined in \cite{Belczynski2008} which delineate whether mass transfer remains dynamically stable or enters a common envelope (CE) which dramatically shrinks the orbit of the BH-LC progenitor. For stable mass transfer our treatment follows the treatment described in \citet{Hurley2002}. We assume that the donor loses mass with a rate that steeply increases with the amount that the donor radius overfills its Roche radius. We assume that the accretor is able to accept mass at $10$ times its thermally limited rate (a star's mass divided by its thermal time) during the main sequence, Hertzsprung gap, and core helium burning phase and at an unlimited rate during any giant-like phases. All mass that is not accreted is assumed to leave the binary with the specific angular momentum of the accretor. 

\cosmic\ employs the $\alpha\lambda$ prescription for CE evolution where $\lambda$ is a parameter which varies the binding energy of the donor's envelope based on its structure and $\alpha$ defines how efficiently orbital energy is used to eject the envelope \citep{Paczynski1976, Livio1984, Tout1997}. We set $\alpha = 1$ which assumes that the initial orbital energy is converted into ejecting the envelope with $100\%$ efficiency. We set the binding energy parameter, $\lambda$, as defined in the Appendix of \cite{Claeys2014}. We assume the default choice in \cite{Claeys2014} such that no recombination energy from ionization in the donor envelope is considered in the envelope ejection \citep[recombination contributes only a few percent to the overall energy budget for massive stars;][]{Kruckow2016}. All stable RLOF and CE interactions are assumed to perfectly circularize the binary.

The distribution of birth kicks BHs receive is quite uncertain because of the limited numbers of detected stellar BHs \citep[e.g.,][]{Repetto2012,Repetto2015,Atri2019} and the complexity of modeling SN explosions from first principles \citep[e.g.,][]{Woosley2002,Sukhbold2018,Pejcha2020}.
For this study, we consider {\bf the two most} widely used mechanisms for the formation of the compact objects: the ``rapid" and ``delayed" mechanisms presented in \citet{Fryer2012}. The two mechanisms differ in the growth time of instabilities in the convective region outside of the proto-compact object which re-initiate the post core-bounce shock to complete the explosion. The primary difference between BH populations formed with the rapid and delayed mechanisms is the existence (rapid) or lack of (delayed) a mass gap separating neutron stars (NS) and BHs between $\sim3\,\rm{M_{\odot}}$ and $\sim5\,\rm{M_{\odot}}$. Each prescription includes a recipe for the partial fallback ($f_{\rm{fb}}$) of the stellar envelope. In addition to its effect on the BH masses, SN fallback reduces the strength of the natal kicks by a factor of $1-f_{\rm{fb}}$ from the kick strength drawn randomly from a Maxwellian distribution with $\sigma = 265\rm{km/s}$ \citep{Hobbs2005} and distributed isotropically.  Throughout this work we test both models and call our simulated populations \rapid\ and \delayed\ after the name of the SN prescriptions used to create them.  

\subsection{Convergence of present-day binary properties}
\label{S:convergence}
COSMIC is designed to adaptively choose the size of a simulated population based on a convergence criteria that tracks how the shape of parameter distributions for binaries, such as mass, orbital period ($\porb$), and eccentricity ($Ecc$) change as the population grows. This is done by iteratively simulating populations of size $N_{\rm{sim}}$ and adding each newly simulated population to the total population on each iteration. Histograms of the binary parameter data are also made at each iteration and compared bin by bin using a criterion inspired by matched filtering techniques \citep[e.g. Eq. 6 of ][]{Chatziioannou2017} and defined as
\begin{equation}
   match = \frac{\sum_{k=1}^{N}P_{k,i}P_{k,i+1}}{\sqrt{\sum_{k=1}^{N}P_{k,i}P_{k,i}\sum_{k=1}^{N}P_{k,i+1}P_{k,i+1}}},
\end{equation}
where $P_{k,i}$ represents the height of bin $k$ on the $i\rm{th}$ iteration \citep{Breivik2020}. The match criteria approaches unity as the shapes of each histogram converge. We use Knuth's Rule to determine the binwidths of each binary parameter histogram for the $i+1$ simulation iteration\citep{Knuth2019}. At each iteration we record the total amount of ZAMS mass drawn in single and binary stars to scale the converged population up to a Galactic population. We continue to simulate binaries until $match > 1-10^{-5}$ for the BH mass ($\mbh$), LC mass ($\mlc$), $\porb$, and $Ecc$. 

\subsection{Creating synthetic Milky-Way population}
\label{S:galactic-realisation}
We generate synthetic Milky-Way populations by sampling with replacement from the converged population of simulated binaries. To determine the number of BH-LC binaries in the Milky Way at present, we scale the number of BH-LC binaries in the converged population as
\begin{equation}
   N_{\rm{BH-LC,MW}} = N_{\rm{BH-LC,sim}}\frac{M_{\rm{\bf m12i}}}{M_{\rm{sim}}},
\end{equation}
where $M_{\rm{\bf m12i}}$ is the total amount of mass formed in galaxy \textbf{m12i}, and $M_{\rm{sim}}$ is the mass of single and binary stars drawn to produce the simulated BH-LC population.  We assign each of the $N_{\rm{BH-LC,MW}}$ BH-LCs a complete set of stellar and orbital parameters including primary mass, secondary mass, age, metallicity, eccentricity, $\rm{P_{orb}}$, and luminosity from the converged population according to the unique binary id. The sampled binaries are then assigned a Galactocentric position by minimizing the difference in age and metallicity between the binary and star particles in the \textbf{m12i} galaxy. Since multiple BH-LCs can be assigned to a single star particle, the BH-LCs are distributed in a sphere centered around the star particle, where the radius of the sphere is determined using an Epanechnikov kernel with kernel size inversely proportional to the local density \citep{Sanderson2020}. Each BH-LC is also assigned an orientation with respect to the line-of-sight from Earth by sampling the inclination ($i$) uniformly in $\cos(i)$, and argument of periapsis ($\omega$) and longitude of ascending node ($\Omega$) uniformly between $0$ and $2\pi$. 

The number of BH-LCs resolvable by \gaia\ (described in \autoref{S:detectability}) varies depending on the random assignment of the parameters described above. To incorporate the effects of this variance, we repeat the process described above to generate $200$ realisations of the Milky-Way population for each of the \rapid\ and \delayed\ models, producing $400$ Milky Way realizations in total.

\section{Detectability of BH-LC binaries using \gaia}
\label{S:detectability}
In this section we describe how we consider several important aspects that determine detectability of a BH-LC by \gaia, including the size of the sky-projected orbit of the LC with respect to the astrometric precision of \gaia, the number of \gaia\ observations at each sky position, and the effects of extinction and reddening on the target LC population. 

\subsection{Reddening}
\label{S:reddening}
Extinction and reddening from interstellar dust have a significant effect on the observation of stars distributed in the Galactic disk. To account for the extinction due to interstellar dust we apply an extinction correction to all LCs in our population. We determine the extinction corrections using a complete three dimensional position-dependent dust map which combines three different dust maps \citep{Drimmel2003,Marshall2006,Green2019} covering different regions of the Milky Way as implemented in \dustmaps\ \citep{Bovy2016}.\footnote{\citet{Green2019} includes data from \gaia\ DR2, Pan-STARRS 1, and 2MASS. \citet{Marshall2006} is based on only the 2MASS data. \citet{Drimmel2003} takes the data from COBE/DIRBE NIR observations.}  

\subsection{Resolving the sky-projected LC orbits by \gaia}
\label{S:resolve}
For a BH-LC binary orbit to be resolved by \gaia\ we require that the extinction-corrected LC magnitude be brighter than \gaia's limit of $G<20$. \gaia\ can observe stars fainter than $G=21$, but in practice these are too faint to detect astrometric motion due to binarity. We further require that the projected angular size of the binary, $\alpha \geq \sigma_{\xi}$ (``optimistic") and $\alpha \geq 3\times \sigma_{\xi}$ (``pessimistic"), where $\sigma_{\xi}$ is \gaia's single position pointing error \citep{Lindegren2018}. In addition to the criteria for resolving the LC's orbital motion, we impose the condition $\porb\leq10\,\yr$ to ensure that each binary completes a full orbit within the duration of \gaia's extended mission. 

Since every BH-LC has a unique position and orientation in the Galaxy, we determine the projected angular size of the binary on the sky using the Thiele-Innes constants:
\begin{eqnarray} \label{eq:TIelements} 
\mathcal{A} &= a_{\LC}(\cos\  \omega\  \cos\  \Omega -  \sin\  \omega\  \sin\  \Omega\  \cos\ i)\\
\mathcal{B} &=  a_{\LC}(\cos\  \omega\  \sin\  \Omega\ +  \sin\  \omega\  \cos\  \Omega\  \cos\ i)\\
\mathcal{F} &=  a_{\LC}(-\sin\ \omega\ \cos\ \Omega\ -  \cos\ \omega\ \sin\ \Omega\ \cos\ i)\\
\mathcal{G} &=  a_{\LC}(-\sin\  \omega\  \sin\ \Omega\ +  \cos\ \omega\ \cos\ \Omega\ \cos\ i).
\end{eqnarray} 
In the equations above, $a_{\rm{LC}}$ is the semimajor axis of the LC orbit defined as 
\begin{equation}
    a_{\LC}=a\frac{\mbh}{\mlc+\mbh},
\end{equation}
\noindent where $a$ is the binary semimajor axis and $\mlc$ ($\mbh$) is the mass of the LC (BH). The projected cartesian position of the LC is then defined by
\begin{eqnarray}
x_{\rm{project, LC}} &= \mathcal{AX}+\mathcal{FY} \label{eq: xTrans}\\
y_{\rm{project, LC}} &= \mathcal{BX}+\mathcal{GY} \label{eq: yTrans}, 
\end{eqnarray}
where, $\mathcal{X}=\cos\,E-Ecc$, $\mathcal{Y} =\sqrt{1-Ecc^2}\sin\,E$, and $E$ is the eccentric anomaly. Finally, we define the projected size of the LC orbit, $a_{\rm{project, LC}}$, on the sky in terms of the semimajor axis of the ellipse defined by \autoref{eq: xTrans} and \ref{eq: yTrans}, as
\begin{equation}
    \alpha = \frac{a_{\rm{project, LC}}}{D},
\end{equation}
\noindent where $D$ is the distance to the binary.

\subsection{Astrometric mass measurement}
\label{S:mass-constraint}
The most promising way of discovering BH-LC binary candidates is through astrometric mass estimation of the BH, which does not suffer the inclination degeneracy present in radial velocity (RV) measurements \citep[e.g., see discussion in][]{Andrews_2019}. If the mass of the unseen companion to an LC is confirmed to be larger than the minimum mass of a BH then the unseen companion is likely a BH candidate. We showed in \citet{Andrews_2019} that the astrometric mass function error 
can be approximated as
\begin{equation}
\label{eq:mf-fractional}
   \left|\frac{\Delta(M_{\rm{BH}}^{3}\mtot^{-2})}{M_{\rm{BH}}^{3}\mtot^{-2}}\right| = 0.9\Big(\frac{\sigma_{\xi}}{\alpha}\Big)\Big(\frac{N}{75}\Big)^{-1/2}
\end{equation}
\noindent where $\mtot = \mbh + \mlc$ and $N$ is the number of times a binary is observed by \gaia. 

We determine the position-dependent $N$ for each BH-LC in the following way. We divide the sky into $164,838$ grids of equal angular area on the sky-plane. We take 360 equal intervals in declination, each $0.5^\circ$ apart between $-90^\circ$ to $90^\circ$. We divide RA for each of the above dec into $n_{\rm{RA}}$ bins where $n_{\rm{RA}}$ is the nearest integer $\geq720\times\cos(\rm{dec})$. Our grid points ensure that the angular sky-projected area between any two adjacent grid points in RA and dec is smaller than the field-of-view ($0.72^\circ\times0.69^\circ$) for each of \gaia's telescopes \citep[e.g.,][]{Lindegren2012}. Using the \gaia\ Observation Forecast Tool (GOST)\footnote{https://gaia.esac.esa.int/gost/} we calculate and store the number of times \gaia\ would observe each grid point in five years ($N_5$; between 2014-19). We find the value of $N_5$ for any BH-LC binary in our Milky-Way realisations, simply by looking up this number corresponding to the nearest grid point from the RA and dec of that particular model binary. Finally, we set $N=2N_5$ to account for a $10\,\rm{yr}$ observation duration.

If the LC mass is estimated using the astrometrically derived distance, then this mass measurement can be combined with the mass function measurement described above to determine the BH mass measurement accuracy $\Delta\mbh$. We use error propagation based on \autoref{eq:mf-fractional} to estimate $\Delta\mbh$\footnote{Derivation of \autoref{eq:delta-mbh} is shown in the Appendix}: 
\begin{eqnarray}
    \label{eq:delta-mbh}
    \left|\frac{\Delta\mbh}{\mbh}\right| & = & 0.1
    \left|\frac{\Delta\mlc}{0.1\mlc}\right| \left(\frac{2\mlc}{\mtot+2\mlc}\right) \nonumber\\
    & + & \left|\frac{\Delta(M_{\rm{BH}}^{3}M_{\rm{tot}}^{-2})}{M_{\rm{BH}}^{3}M_{\rm{tot}}^{-2}}\right|\left(\frac{\mtot}{\mtot+2\mlc}\right) \nonumber\\
    & = & 0.1
    \left|\frac{\Delta\mlc}{0.1\mlc}\right| \left(\frac{2\mlc}{\mbh+3\mlc}\right) \\
    & + & 0.9 \left(\frac{\sigma_\xi}{\alpha}\right) \left(\frac{2N_5}{75}\right)^{-1/2}\left(\frac{\mbh+\mlc}{\mbh+3\mlc}\right)\nonumber
\end{eqnarray}

To determine the number of potential BH candidates, we assume that all LCs have mass measurements to $10\%$ accuracy. We select BH candidates from our simulated population if $\mbh>\mlc$ and the BH has a mass measurement such that $\mbh - \Delta \mbh \geq 3\,\msun$. We impose the former condition to eliminate tricky scenarios where confirming a BH candidate is difficult because of the possibility that the LC may dominate the total light not because its companion is a dark remnant, but because it is a lower-mass and hence fainter star.    

\section{RESULTS}
\label{S:results}
In this section we describe the BH-LC populations, both for the entire Milky Way and the subset resolvable by \gaia. We include both the overall detection rate as well as the stellar and orbital characteristics of the BH-LC binaries. 
\subsection{Present-day properties of simulated BH-LC binaries}
\label{S:fixedpop}
In order to gain insight for the underlying population of BH-LCs in the Milky Way at present, we first focus on the observable properties of the BH-LC populations created in our models without imposing any observational selection effects.   

\begin{figure}
    \plotone{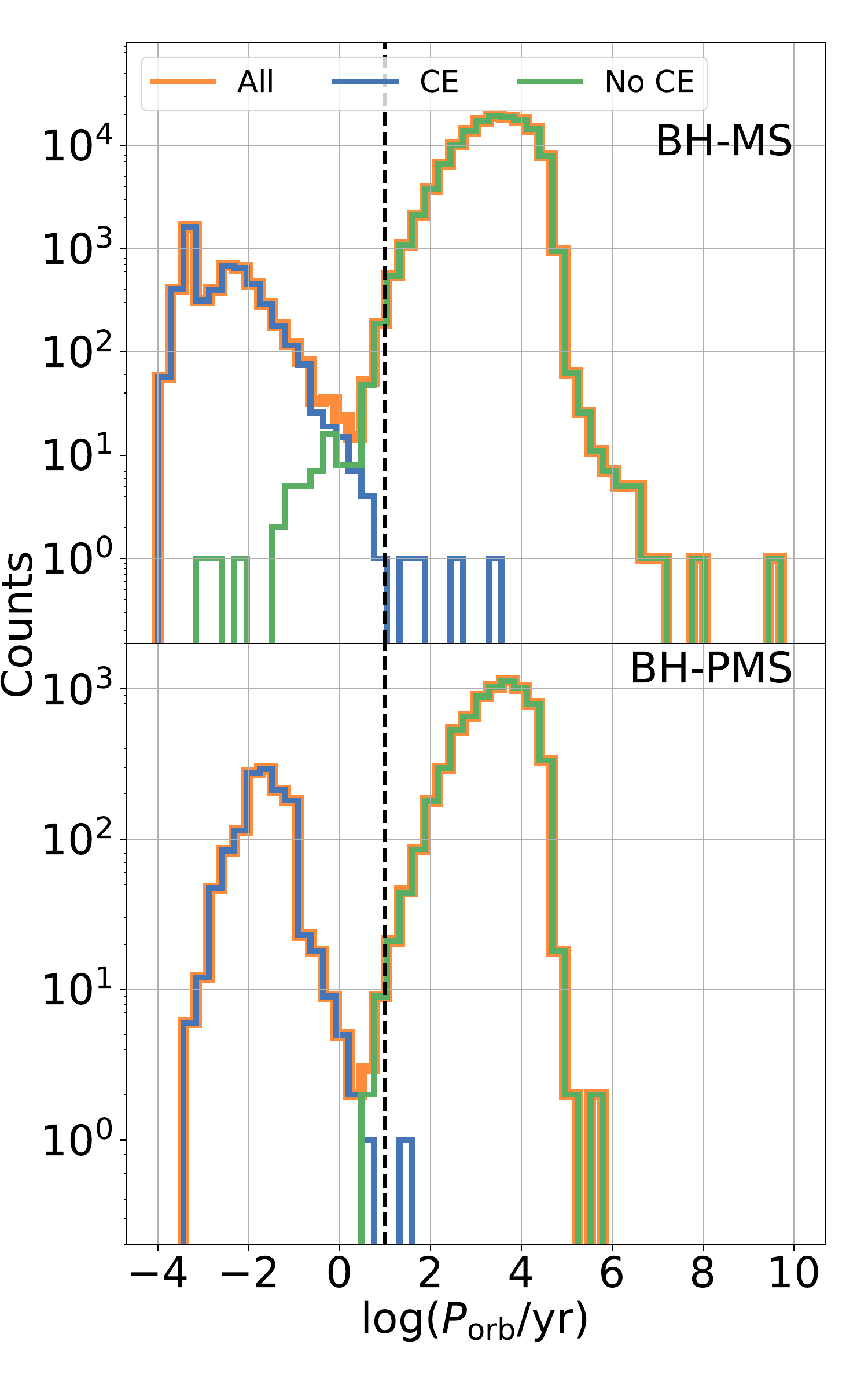}
    \caption{The distribution of orbital periods ($\porb$) for all BH-LC binaries at present time in the Milky Way in our \rapid\ model (\autoref{S:methods}). The top and bottom panels show binaries with MS and PMS companions. Orange, blue, and green histograms denote distributions for all BH-LC binaries, those that went through at least one CE evolution, and those that never went through a CE evolution. The short-period peak is dominated by BH-LC systems created via at least one CE episode, whereas, the long-period peak is created primarily by BH-LC binaries that never went through a CE episode. In our analysis we consider BH-LC binaries with $\porb\leq10\,\yr$ (the vertical black dashed line to guide the eye), such that a $10$-$\yr$ \gaia\ mission can resolve the full orbit. Interestingly, $10$-$\yr$ \gaia\ mission potentially can detect BH-LC binaries created via the CE channel as well as those that never went through a CE evolution. (The equivalent figure for our \delayed\ model is presented in the Appendix.)
    }
    \label{fig:fixedpop-porb}
\end{figure}

\autoref{fig:fixedpop-porb} shows the overall distribution for the orbital periods ($\porb$) of all BH-LC binaries in the Milky Way at present from the \rapid\ model (\autoref{S:methods}). The distribution of $\porb$ for BH-LC binaries in the \delayed\ model is very similar (shown in the appendix). The $\porb$ distributions are clearly bimodal for both BH-MS and BH-PMS binaries for both of the adopted SN models. In both cases, the short-period peak is dominated by binaries created via at least one CE episode, whereas, the long-period peak is created primarily with BH-LC binaries that never undergo a CE. The location of the trough for all cases is between $\porb\sim 1$--$10\,\yr$. 
The majority ($\sim96\%$) of BH-LC binaries with $\porb\leq 10\,\yr$ have gone through at least one CE evolution independent of the adopted SN mechanism. 
Interestingly, the extended $10\,\yr$ mission of \gaia\ allows sampling of the separation between the CE and non-CE peaks. 
While outside the scope of this work, future studies could investigate how the observed shape and width of the trough can constrain uncertain CE physics.

\begin{figure}
    \plotone{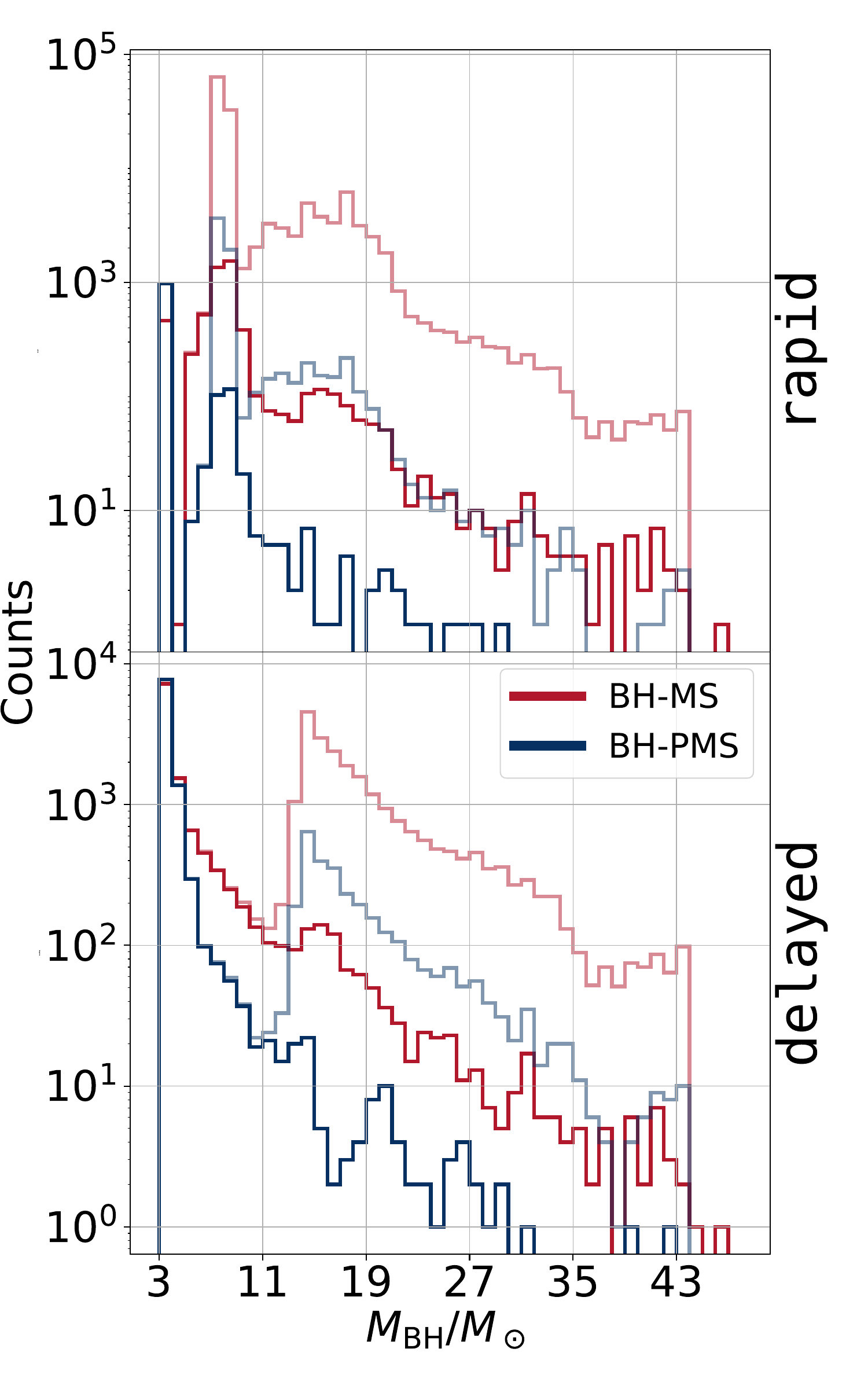}
    \caption{Distribution of BH mass for present-day populations from the \rapid\ (top) and \delayed\ (bottom) models. The blue and red curves represent MS and PMS companions, respectively. The faded and bright lines denote all BH-LC binaries without any constraints on $\porb$ and those satisfying $\porb\leq10\,\yr$. The major difference between the two distributions come in the range $\mbh/\msun=3$---$5$. In case of the \rapid\ model, all BHs within $\mbh/\msun=3$---$5$ come from AIC of NSs; we assume that the limiting mass for NSs is $3\,\msun$ (\autoref{S:methods}). Whereas, for \delayed, BHs formed via core-collapse SN can have masses down to $3\,\msun$. As a consequence, the \delayed\ model consists of a relatively higher fraction [$25\ (71)\%$] of BHs in this mass range of compared to the \rapid\ model [$0.3\ (12)\%$] with MS (PMS) companion. In both cases, $\mbh$ is distributed in a wide range. 
    }
    \label{fig:fixedpop-mbh}
\end{figure}
One of the most prominent differences between the two SN mechanisms manifests in the $\mbh$ distribution; while the rapid prescription for core-collapse SN does not allow production of BHs in the mass range $3$ to $5\,\msun$, (often called the `mass-gap' between neutron stars (NSs) and BHs), the \delayed\ prescription predicts no mass gap.
\autoref{fig:fixedpop-mbh} shows the $\mbh$ distributions for BH-LC binaries at present day for the \rapid\ and \delayed\ models. The distributions are clearly different in the range between $3$ and $5\,\msun$, as expected. {\em All} BHs with $\mbh$ between $3$ and $5\,\msun$ in the \rapid\ model originate from accretion-induced collapse (AIC) of NSs. In contrast, in the \delayed\ model, the AIC BHs contribute at $\sim 15\ (55)\%$ for BHs MS (PMS) companions and the rest come from core-collapse SN.
The $\mbh$ distribution continuous all the way down to $\mbh/\msun=3$, our adopted mass that separates NSs from BHs in the \delayed\ model, whereas, for the \rapid\ model, there is a clear separation between the mass-gap BHs and the rest. 

Interestingly, all BH-LC binaries with $\mbh/\msun\lesssim7$ ($10$) in the \rapid\ (\delayed) model have $\porb\leq10\,\yr$ leading to identical populations for the BH-LC binaries with lower-mass BHs regardless of $\porb$ restrictions. This is because lower-mass BHs receive larger natal kicks in both SN prescriptions we have considered \citep{Fryer2012} while for higher-mass BHs, natal kick magnitudes are reduced depending on the amount of mass that falls back onto the proto- compact object \citep[e.g.,][]{Belczynski2008}. The large natal kicks for low-mass BHs break their progenitor binaries with orbits $\porb/\yr\gtrsim10$, while, the more massive counterparts in wide orbits can remain intact. If \gaia\ is able to characterize BHs in binaries with $\mbh\sim10\,\msun$ and $\porb\leq10\,\yr$, strong constraints can be placed on the strength of natal kicks for these systems.

A few other features, originating from a combination of the adopted initial stellar mass function, metallicity and age distributions, and the complex binary star evolution of the initial population of binaries, are noticeable and different in the two models. For example, in the \rapid\ model a peak in the range $\mbh/\msun\approx7$ to $9$ come from the evolution of metal-rich ($\metal/\metal_\odot\geq 0.75$) systems. Another less prominent and broader peak between $\mbh/\msun\approx13$ and $19$ consists of systems that are relatively metal poor ($\metal/\metal_\odot\leq 0.5$). In contrast, in the \delayed\ model the distribution peaks near $\mbh/\msun=3$ and again near $\mbh/\msun\approx15$. 

\begin{figure}
    \plotone{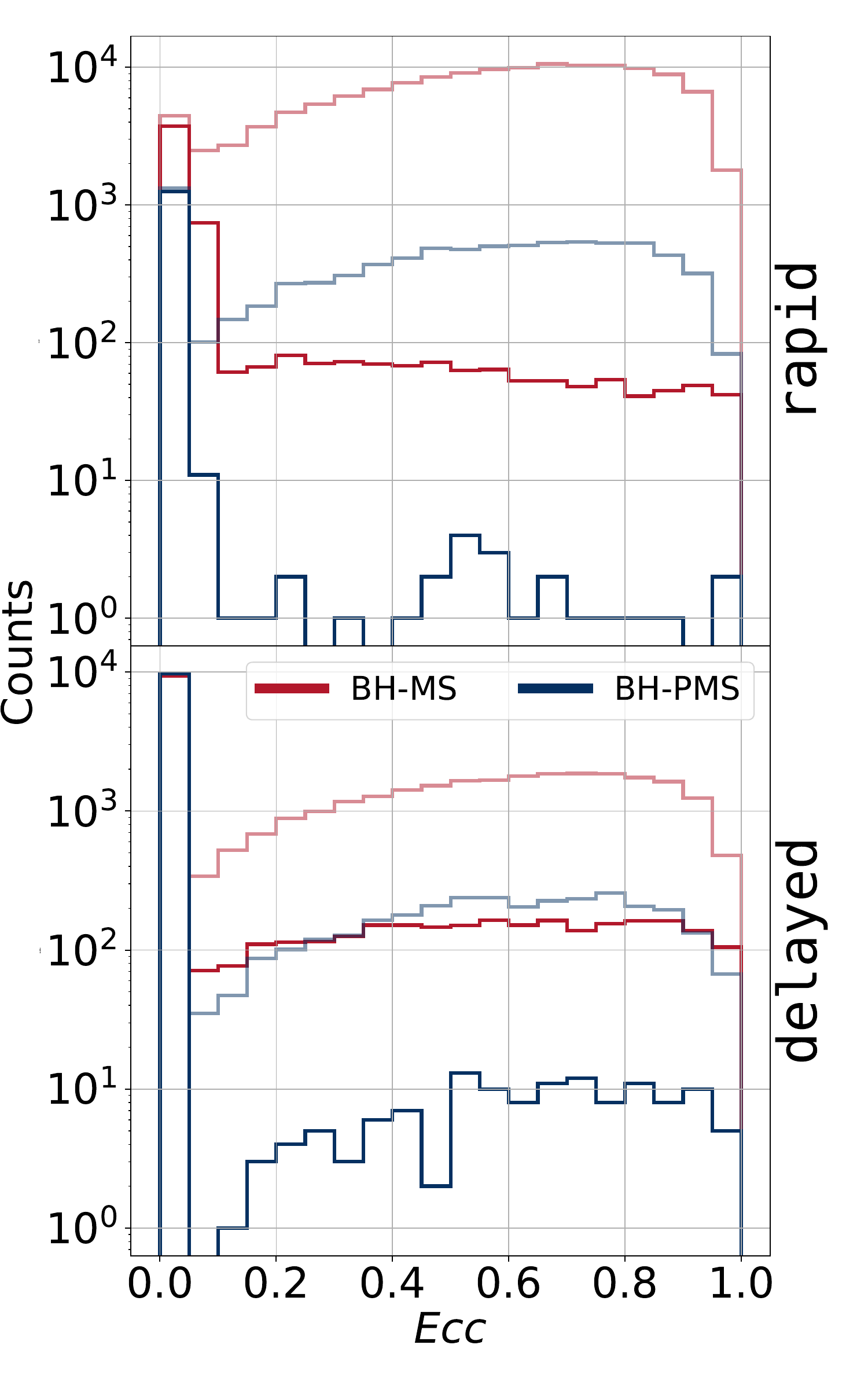}
    \caption{Same as \autoref{fig:fixedpop-mbh} but showing the distribution for the orbital eccentricities ($\ecc$). In all cases there is a prominent peak for near-zero eccentricity, especially for BH-LC binaries with $\porb/\yr\leq10$. This is because a large fraction of short-period BH-LC binaries go through CE evolution and are significantly affected by tides. The BH-MS (BH-PMS) binaries in the \rapid\ and \delayed\ models contain $0.8\%$ ($0.3\%$) and $7\%$ ($1\%$) $\porb/\yr\leq10$ systems with $Ecc\geq0.1$, respectively. The \delayed\ model contains higher fraction of eccentric BH-LC binaries compared to the \rapid\ model. This can be attributed to typically larger natal kicks in case of \delayed. The fraction of near-circular orbits is less in case of BH-MS compared to BH-PMS in both models since tidal circularisation is more effective in BH-PMS binaries than in BH-MS binaries.  
   }
    \label{fig:fixedpop-ecc}
\end{figure}
\autoref{fig:fixedpop-ecc} shows the distributions of BH-LC orbital eccentricities for the \rapid\ and \delayed\ models. We find that about $2\%$ ($15\%$) and $27\%$ ($76\%$) of BH-MS (BH-PMS) binaries with $\porb/\yr\leq10$ have near-circular ($\ecc\leq10^{-2}$) orbits for the \rapid\ and \delayed\ models, respectively. The majority of the BH-LC systems with $\porb\leq10\,\yr$ go through at least one CE evolution stage with the BH-progenitor's envelope which circularizes the binary before BH formation (\autoref{fig:fixedpop-porb}). Moreover, even the small fraction of the short-period BH-LC binaries that never went through a CE phase can be significantly affected by tidal circularisation. In spite of this, the high fraction of eccentric BH-LC binaries primarily result from natal kicks during BH formation. The differences in the explosion mechanism between the rapid and delayed SN prescriptions lead to typically higher natal kicks in the \delayed\ models. As a result, the fractions of short-period BH-MS (BH-PMS) binaries with significant $\ecc\geq0.1$ are $0.8\%$ ($0.3\%$) and 7\% (1\%) in our \rapid\ and \delayed\ models, respectively. The connection between larger eccentricities and stronger natal kicks is a robust prediction that is independent of our chosen SN prescription. For both SN mechanisms, BH-PMS binaries usually have a smaller fraction of eccentric systems compared to BH-MS binaries, since the former have larger LC envelopes which are affected more strongly by tides after BH formation.  

\begin{figure*}
    \plotone{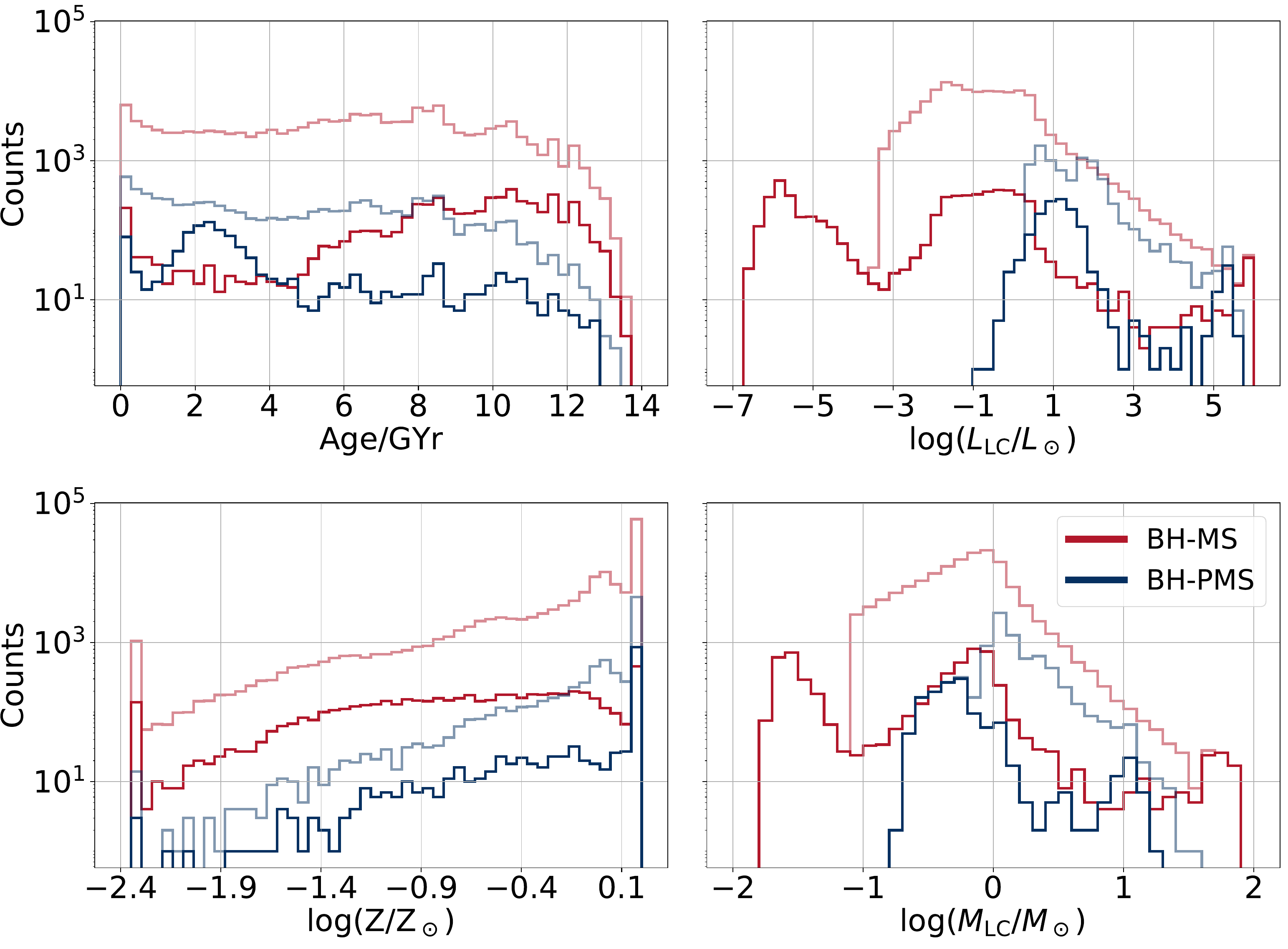}
    \caption{Distributions for present-day age (top-left), luminosity (top-right), metallicity (bottom-left), and mass (bottom-right) of the LC from our \rapid\ model. Red and blue lines denote MS and PMS companions. The faded and bright lines denote all BH-LC binaries and those satisfying $\porb\leq 10\,\yr$, respectively. Simulated present-day BH-LC binaries in the Milky Way show large ranges in age, metallicity, $\mlc$, and $\llc$. (The equivalent figure for our \delayed\ model is presented in the  Appendix.)
    }
    \label{fig:fixedpop-otherprop}
\end{figure*}
\autoref{fig:fixedpop-otherprop} shows the distributions for age, metallicity, mass, and luminosity of the LCs for all BH-LC binaries and those satisfying $\porb\leq10\,\yr$ for the \rapid\ model. The corresponding figure for the \delayed\ model is included in the Appendix. In contrast to the eccentricity and $\mbh$ distributions, the LC properties primarily depend on the stellar IMF, binary stellar evolution physics, and the star formation and metallicity evolution in the Milky Way; they do not strongly depend on the details of the assumptions of core-collapse SN physics. As a result, the distributions of LC properties are very similar in the \rapid\ and \delayed\ models. We find that the ages of the BH-LC progenitors spans several Gyr covering the full range of the star particle ages in the galaxy model \textbf{m12i}. As a consequence, the BH-LC progenitors also have a wide range in metallicities. 
The pile up at the last metallicity bins near $\log(\metal/\metal_\odot)=-2.3$ and $0.2$ is simply because the star particles in the model galaxy \textbf{m12i} have a wider range in $\metal$ relative to the allowed $\metal$ range in COSMIC which is based on the single star fits of BSE \citep[][see also the discussion in \autoref{S:initialising}]{Hurley2000}. 
Nevertheless, the wide range in metallicities of our predicted present-day BH-LC binaries in the Milky Way indicates that if a sizeable population of BHs are detected, metallicity-dependent constraints on the properties of BHs may be obtained. Note that our earlier work made a simplifying assumption that used two fixed metallicity values corresponding to the thin and thick disks of the Milky Way \citep{Breivik_2017}. Using more realistic metallicities that depend on the star-formation history in the Milky Way is one of the key improvements in the present study. 

Almost all low-mass ($\lesssim 0.1\,\msun$) MS companions to the BH binaries show $\porb\leq10\,\yr$. This is likely related to the CE evolution. 
Orbits with lower LC masses (on an average) contain a smaller reservoir of orbital energy to eject the envelope. As a result, BH's with lower-mass LCs tend to have shorter post-CE orbits.
Although, not as prominent, the same trend is observed for BH binaries with PMS companions; below $\mlc\sim1\,\msun$, all BH-PMS binaries have $\porb\leq10\,\yr$. 

A small ($\lesssim 0.1\%$) fraction of BH-MS binaries consist of MS stars in excess of $40\,\msun$. We find that in some cases, an initially high-mass ($M\geq 15\,\msun$) MS star accretes from the BH progenitor through stable mass transfer via RLOF and grow. These stars live longer than normal due to rejuvenation \citep[e.g.,][]{Hurley2002}. Most of these systems are also very young ($\lesssim 9\,\myr$). As expected, the mass range for the PMS companions is narrower compared that for the MS companions because the PMS phase is a shorter-lived stage of evolution. In addition, lower-mass LCs must be sufficiently old to advance off the MS leading to a higher relative number of BH-MS binaries with ages larger than $6\,\rm{Gyr}$ relative to BH-PMS binaries.

We estimate that the Milky Way at present harbors $\sim4\times10^5$ ($\sim8\times10^4$) BH-LC binaries according to the \rapid\ (\delayed) model. Of these, $17,621^{+145}_{-144}$ ($29,616^{+181}_{-189}$) have $\porb/\yr\leq10$. Of the BH-LC binaries with $\porb/\yr\leq10$, $9,184^{+120}_{-133}$ ($7,091^{+113}_{-100}$) BH-LC binaries are in detached configurations in the \rapid\ (\delayed) model (\autoref{tab:detection})\footnote{Throughout the paper, the numbers and the errorbars denote the median and the span between the 10th and 90th percentiles from our Milky Way realisations.}. We focus on the detached BH-LC binaries with $\porb/\yr\leq10$ in the Milky Way for \gaia's ability to resolve the LC's orbits throughout the rest of the paper. 

\subsection{BH-LC binaries in the Milky Way resolvable by \gaia}
\label{S:milkyway}
\begin{deluxetable*}{c|c|ccc|ccc|ccc}
\tablecolumns{11}
\tabletypesize{\scriptsize}
\tablecaption{BH-LC binaries in the Milky Way} 
\tablehead{
 \colhead{Model}\vline & \colhead{Type}\vline &
 \multicolumn{3}{c}{Intrinsic}\vline &
 \multicolumn{6}{c}{Resolvable} \\
 \cline{3-5}
 \cline{6-11}
 \colhead{}\vline & \colhead{}\vline & 
 \colhead{All} & \multicolumn{2}{c}{$\porb/\yr<10$}\vline &
 \multicolumn{3}{c}{Optimistic} & \multicolumn{3}{c}{Pessimistic} \\
 \cline{4-11}
 \colhead{}\vline & \colhead{}\vline & \colhead{} &
 \colhead{All} & \colhead{Detached}\vline & 
 \colhead{Resolved} & \colhead{$A_v$-corrected} & \colhead{BH candidate} & \colhead{Resolved} & \colhead{$A_v$-corrected} & \colhead{BH candidates}
 }
\startdata
  & MS & $369,199$ & $14,679^{+142}_{-145}$ & $8,658^{+113}_{-123}$ & $383^{+33}_{-28}$ & $255^{+19}_{-26}$ & $76^{+9}_{-11}$ & $191^{+20}_{-19}$ & $139^{+12}_{-18}$ & $35^{+8}_{-7}$ \\
  \rapid\ & PMS & $18,990$ & $2,947^{+52}_{-71}$ & $529^{+26}_{-27}$ & $88^{+11}_{-12}$ & $36^{+9}_{-7}$ & $31^{+8}_{-6}$ & $47^{+9}_{-9}$ & $14^{+5}_{-4}$ & $12^{+4}_{-4}$ \\
  & total & $388,189$ & $17,621^{+145}_{-144}$ & $9,184^{+120}_{-133}$ & $470^{+32}_{-30}$ & $292^{+21}_{-26}$ & $107^{+11}_{-12}$ & $239^{+21}_{-20}$ & $152^{+14}_{-17}$ & $47^{+9}_{-8}$ \\
   \hline
  & MS & $70,694$ & $24,731^{+166}_{-175}$ & $6,653^{+106}_{-104}$ & $125^{+15}_{-12}$ & $68^{+11}_{-9}$ & $13^{+5}_{-4}$ & $40^{+8}_{-8}$ & $26^{+7}_{-6}$ & $6^{+3}_{-3}$ \\
  \delayed\ & PMS & $6,355$ & $4,886^{+47}_{-48}$ & $443^{+26}_{-28}$ & $57^{+10}_{-8}$ & $27^{+6}_{-7}$ & $11^{+5}_{-3}$ & $19^{+5}_{-5}$ & $9^{+3}_{-4}$ & $4^{+4}_{-2}$ \\
  & total & $77,049$ & $29,616^{+181}_{-189}$ & $7,091^{+113}_{-100}$ & $184^{+15}_{-14}$ & $95^{+12}_{-12}$ & $24^{+7}_{-5}$ & $60^{+8}_{-12}$ & $36^{+7}_{-8}$ & $10^{+4}_{-4}$ \\  
\enddata
\tablecomments{BH-LC numbers in the Milky Way predicted in our models. The numbers and errors denote the median and the spread between the $10$th and $90$th percentiles across the Milky-Way realisations. ``Intrinsic" denotes the present-day population of BH-LC binaries in the Milky Way. ``Resolvable" denotes the subset of BH-LC binaries resolvable by \gaia's astrometry over a $10$ $\yr$ mission (\autoref{S:resolve}). Any detached BH-LC binary with $\porb/\yr\leq10$ and resolved LC orbit is denoted as ``Resolved." ``$A_v$-corrected" denotes the expected number of resolved BH-LC binaries after correcting for extinction and reddening from interstellar dust (\autoref{S:reddening}). A BH-LC is a ``BH candidate" if \gaia's astrometry can constrain the BH's mass $\mbh-\Delta\mbh\geq3\,\msun$ and $\mbh\geq\mlc$ within the $A_v$-corrected resolved population (\autoref{S:mass-constraint}).
}
\label{tab:detection}
\end{deluxetable*}

In this section we introduce the realistic Milky-Way populations each consistent with the complex and interdependent age-metallicity-Galactic position observed in the Galaxy (\autoref{S:galactic-realisation}).
These realisations are different from each other in the random orientations, the distance, and the Galactic locations of the BH-LC binaries. As a result there are differences in the number of detections by \gaia\ in each realisation. 
These differences provide an estimate of the level of statistical fluctuations in the expected BH-LC numbers resolvable by \gaia. 

\begin{figure}
    \plotone{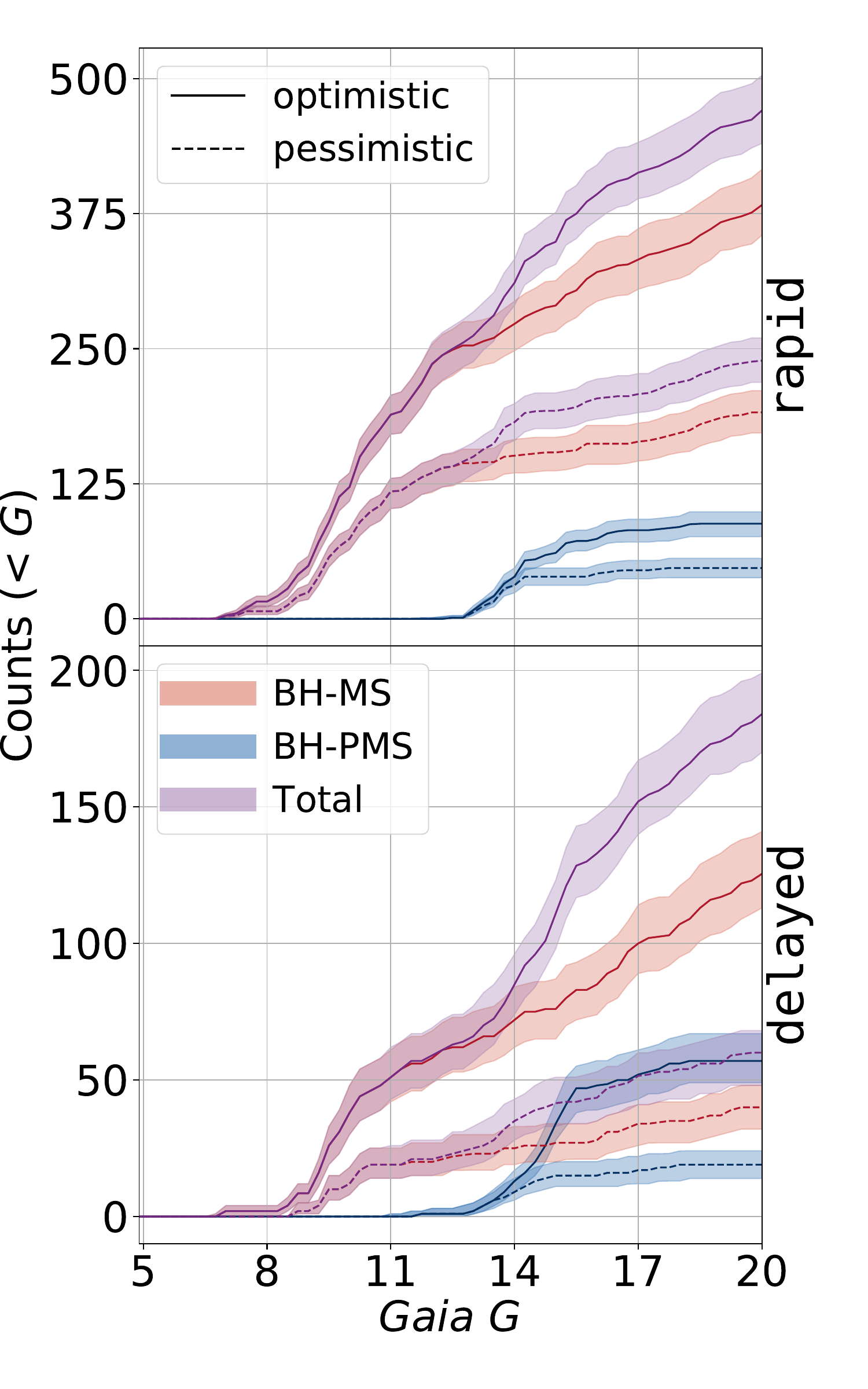}
    \caption{The cumulative number of expected detections of BH-LC binaries by \gaia\ as a function of \gaia's $G$ magnitude for the \rapid\ (top) and \delayed\ models. Purple, blue, and red denote all BH-LC binaries, BHs with PMS, and MS companions, respectively. Lines and shaded regions around them represent the median and the range between the $10$th and $90$th percentiles. The statistical fluctuations come from the 200 realisations of the Milky Way we generate (\autoref{S:galactic-realisation}). Solid and dashed lines denote two different astrometric cuts, $\alpha > \sigma_{\xi}$ (optimistic) and $\alpha > 3\sigma_{\xi}$ (pessimistic), respectively. }
    \label{fig:resolvable-numbers}
\end{figure}

\autoref{fig:resolvable-numbers} shows the cumulative number of BH-LC binaries for which \gaia\ can resolve the motion of the LC in orbit around the BH as a function of \gaia's $G$ magnitude. 
The estimated total number of resolvable BH-LC binaries for the \rapid\ model is between $\approx240$ (pessimistic) and $470$ (optimistic). In case of the \delayed\ model, the expected yield is lower, between about $60$ (pessimistic) and $185$ (optimistic) BH-LCs. The delayed SN prescription typically produces lower-mass BHs and higher natal kick magnitudes compared to the rapid SN prescription. As a result, a higher fraction of the progenitors of BH-LC binaries disrupt in our \delayed\ model. 

\begin{figure*}
    \plotone{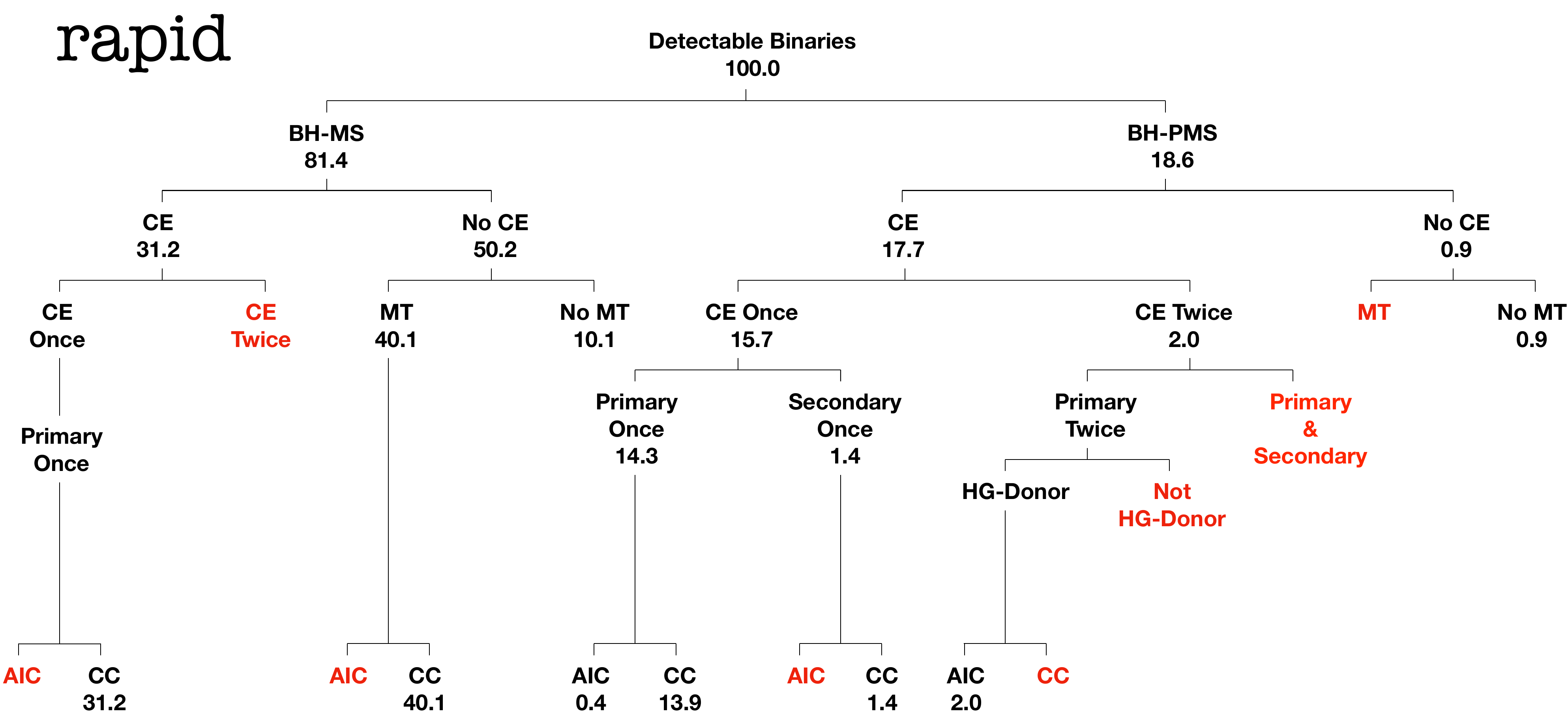}
    \caption{Branching ratios of the different formation channels for the resolvable BH-LC binaries in the \rapid\ model. The branches in red denote no contribution from that channel. CE, MT, CC, and AIC denote common-envelope, stable mass transfer via RLOF, core-collapse SN, and accretion-induced collapse of a NS, respectively. We divide the CE channels based on the number of CE phases the binaries go through as well as the the binary component that initiates the CE. The fractional contribution from each channel is denoted by numbers in percent (rounded to one significant digit after the decimal). Most BH-LC binaries are BH-MS. We further highlight channels created via a HG donor, we allow such binaries to survive in our simulations. Note that BH-LC binaries that ever had a HG donor contribute towards BH-PMS at the level of about $2\%$. }
    \label{fig:resolvable-rapid-branching}
\end{figure*}
We now embark on an investigation of the relative importance of various formation channels of the resolvable BH-LC binaries. Identification of the relative importance of different channels can inform which binary evolution physics of any particular channel matters most in estimating the model populations. In addition, different formation channels can create BH-LC binaries with significantly different properties including $\porb$ and $\ecc$ (\autoref{fig:fixedpop-porb}, \ref{fig:fixedpop-ecc}). Moreover, future BH-LC detections from \gaia\ can inform the relative abundances of various types of BH-LC binaries and shed light on the uncertain aspects of binary stellar evolution including the CE, and BH formation physics.

\begin{figure*}
    \plotone{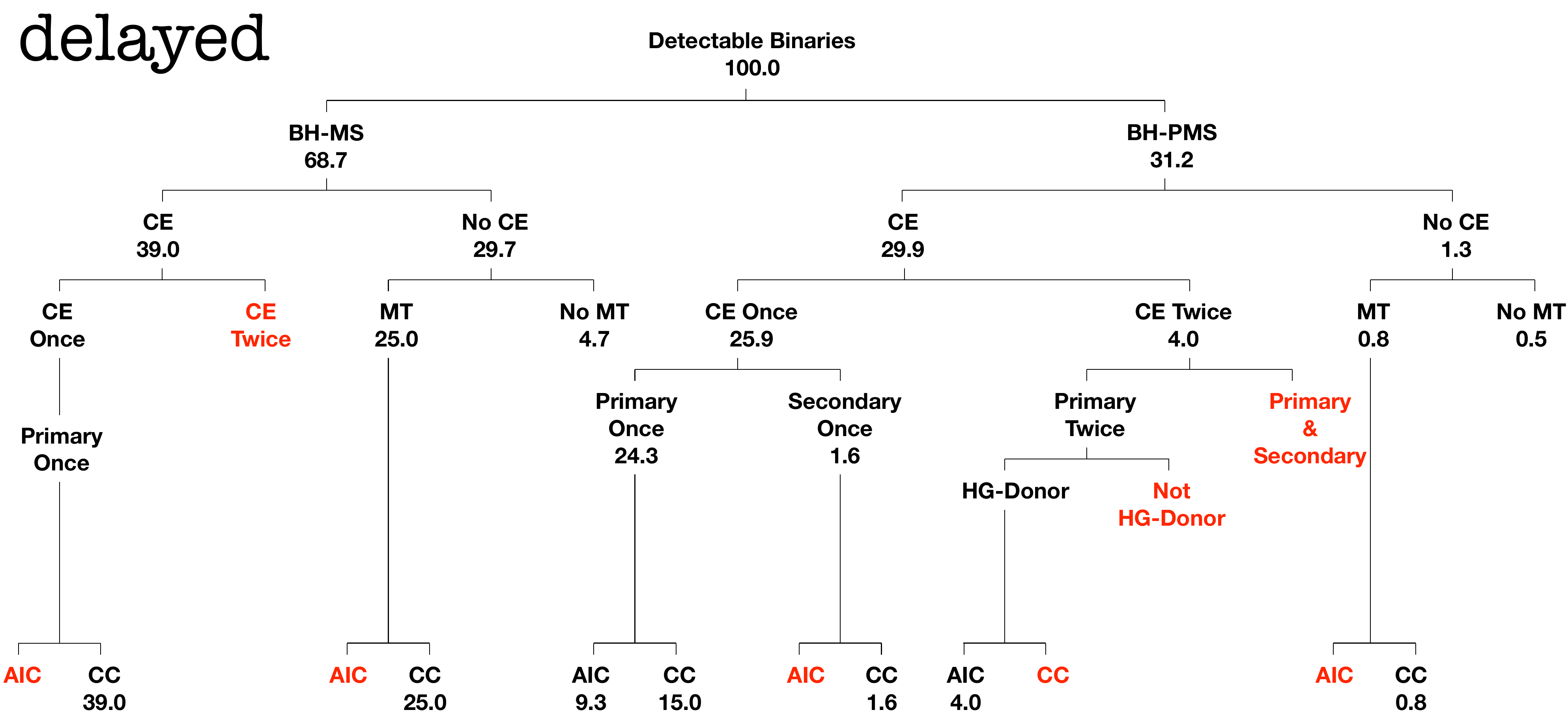}
    \caption{Same as \autoref{fig:resolvable-rapid-branching}, but for the \delayed\ model. BH-MS and BH-PMS created via CE evolution have similar contributions. BH-LCs that at some point had a HG donor contributes at the level of about $4\%$. } 
    \label{fig:resolvable-delayed-branching}
\end{figure*}

\autoref{fig:resolvable-rapid-branching} and \ref{fig:resolvable-delayed-branching} show the detailed evolutionary pathways for the \gaia-resolvable BH-LC binaries and their relative importance. Below we mention the key findings. In case of the \rapid\ model, above $80\%$ of resolvable BH-LC binaries are expected to contain a MS star as a companion. Overall, $\sim50\%$ of all resolvable BH-LC systems come via CE evolution. Note that, the contribution from CE evolution in the intrinsic BH-LC population with $\porb/\yr\leq10$ is much higher ($\sim96\%$; \autoref{S:fixedpop}). 
However, \gaia's astrometry can resolve the typically large orbits of the non-CE BH-LC binaries to larger distances. This selection bias increases the fraction of non-CE BH-LCs in the resolvable population. Looking at the resolvable BH-MS and BH-PMS populations separately, about $40\%$ ($95\%$) of the resolvable BH-MS (BH-PMS) binaries are expected to have undergone at least one CE phase. 

Only about $2.4\%$ of the resolvable BH-LCs contain BHs formed via AIC and all of them have a PMS companion. This is significant in two ways. First, if the mass-gap is real, it will be easily discovered from the resolvable BH-LC population without any significant contamination from BHs created via AIC of NSs. Second, the resolvable BHs in the mass gap are expected to have PMS companions only. 

We find that in about $2\%$ of the resolvable BH-LCs the BH progenitor is the donor while crossing the Hertzsprung Gap (HG) during one of the CE episodes. Whether a binary can survive a phase of CE evolution with a HG donor is uncertain\citep[e.g.,][]{Ivanova2004, Belczynski2008}. In spite of the uncertainty of this channel, we do not exclude these systems from our analysis for the sake of completeness. Interestingly, if we do exclude systems that underwent a CE initiated by a HG donor, then the fraction of resolvable BH-LCs containing mass-gap BHs reduces to only $0.4\%$. 

In the \delayed\ model, the relative importance of the formation channels are somewhat different. MS companions still dominate ($69\%$) the population of resolvable BH-LCs. About $70\%$ of the resolvable BH-LCs form via at least one CE episode and about $13\%$ of the BH-LCs contain BHs created via AIC. 

The extension of \gaia's mission from $5$ to $10$ years allows \gaia\ to target BH-LCs up to $\porb=10\,\yr$ and still resolve the full orbit. According to our \rapid\ (\delayed) model, the total expected yield of resolved BH-LC binaries increases by $34\%$ ($16\%$) simply because of the mission extension and this increase in overall yield almost entirely comes from the increase in detectability of longer-period BH-LC binaries that do not go through CE evolution. We find that \gaia's $10\,\yr$ mission would identify significant fractions ($50\%$ and $30\%$ for \rapid\ and \delayed\ models) of resolvable BH-LC binaries that do not come from the CE evolution channel. If such yields are realised, it may be possible to distinguish resolved BH-LC systems that form via CE evolution from those that do not simply using the $\porb$ distribution.   

\subsection{Key Properties of resolvable BH-LC binaries}
\label{S:milkyway-properties}
\begin{figure*}
    \plotone{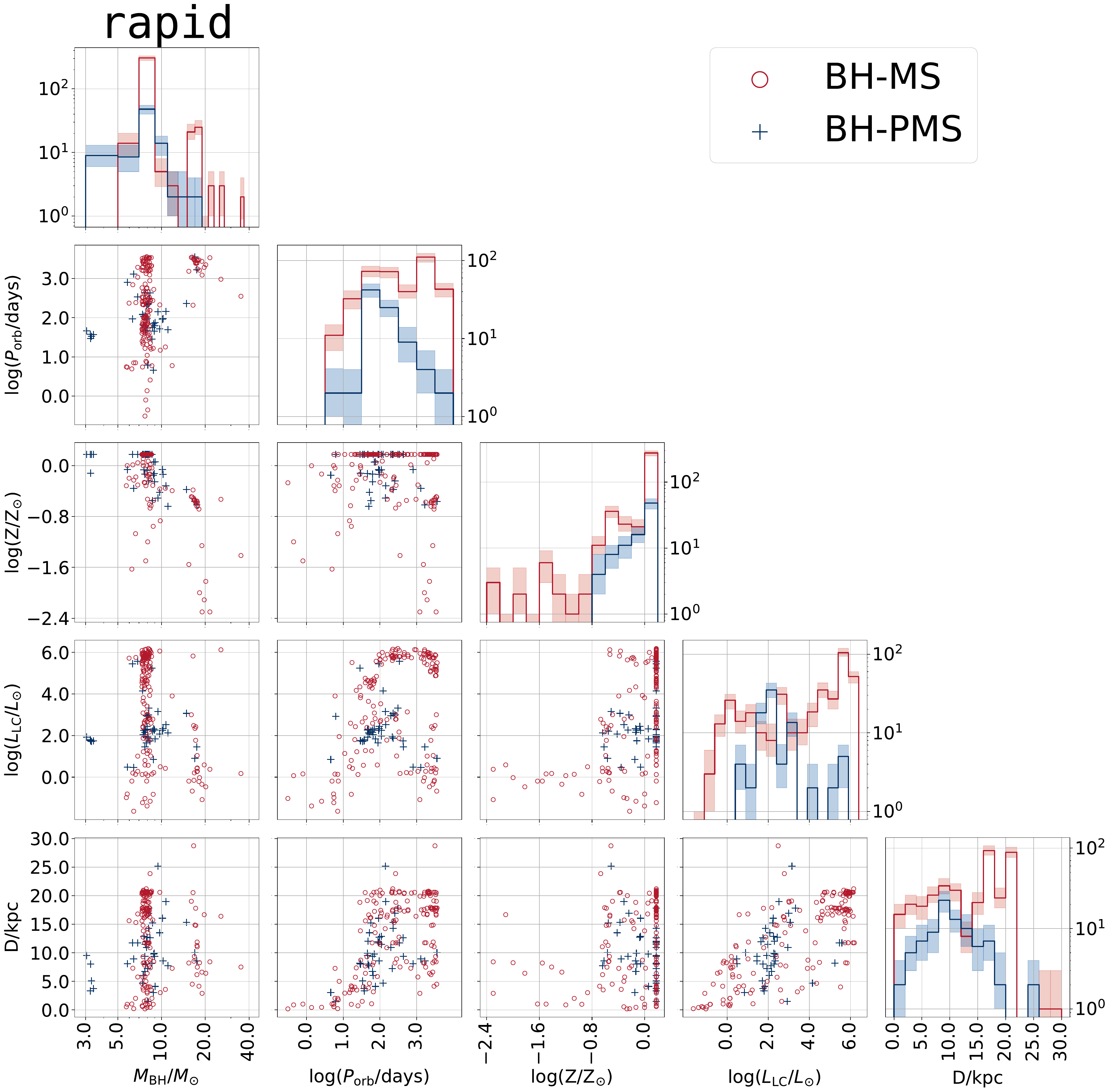}
    \caption{Corner plot showing the observable properties- $\llc$, $D$, $\metal$, $\porb$, and $\mbh$- of BH-MS (red) and BH-PMS (blue) binaries resolvable by \gaia\ in our \rapid\ model. The scatter plots show each unique resolvable binary across all of our Milky Way realisations. Histograms denote the PDF for each property. The shaded regions in the histograms have the same meaning as in \autoref{fig:resolvable-numbers}.  
    }
    \label{fig:resolvable-corner-rapid}
\end{figure*}
\begin{figure*}
    \plotone{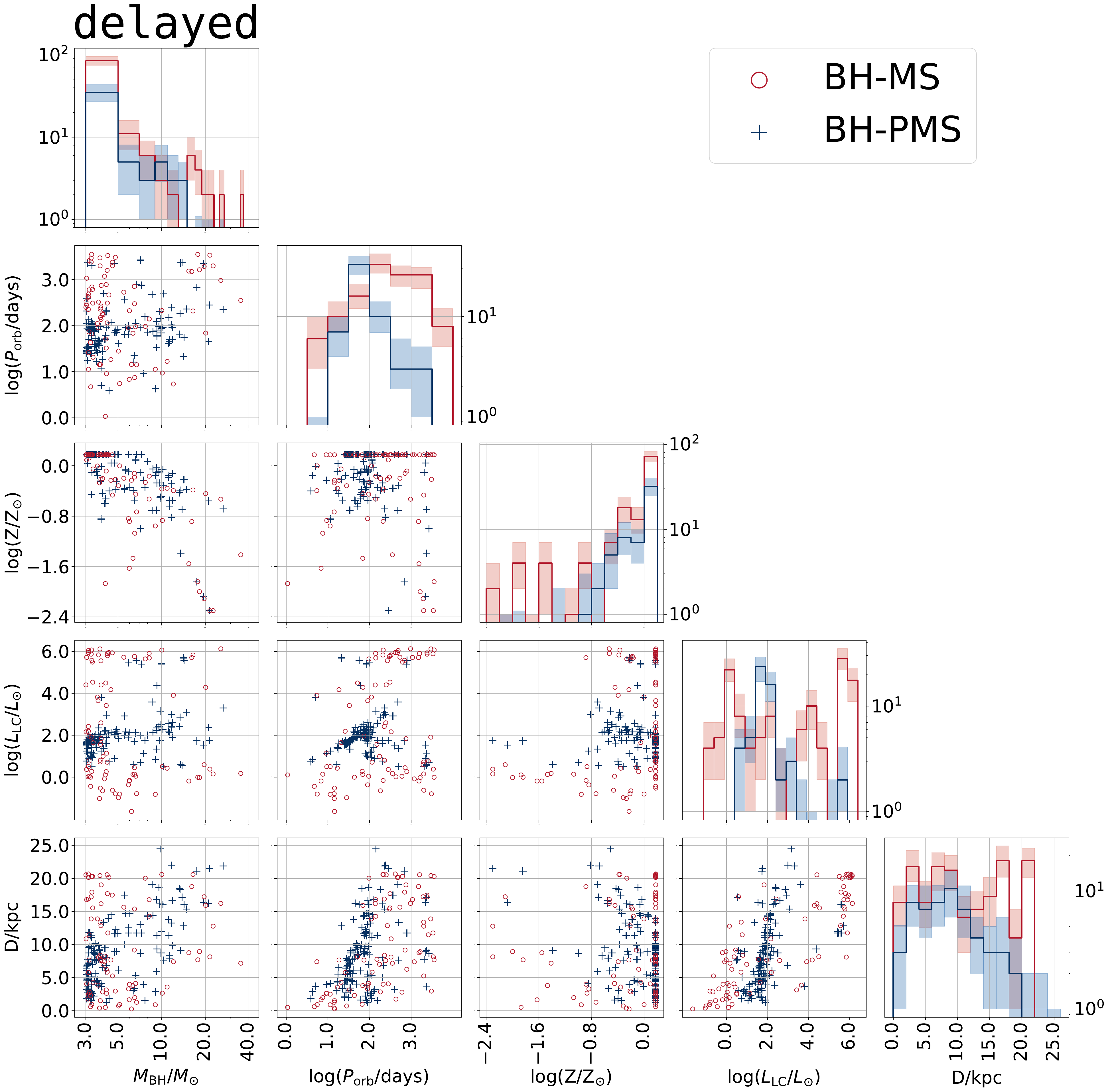}
    \caption{Same as \autoref{fig:resolvable-corner-rapid} but for the \delayed\ model. 
    }
    \label{fig:resolvable-corner-delayed}
\end{figure*}
In this section we highlight some key properties of the resolvable BH-LC binaries. Comparisons between the distribution of properties presented here with those presented in \autoref{S:fixedpop} sheds light on the selection biases of \gaia. \autoref{fig:resolvable-corner-rapid} and \ref{fig:resolvable-corner-delayed} show the corner plots for all resolvable binaries across all Milky Way realisations from our \rapid\ and \delayed\ models. Note that different Milky Way realisations can sample the same binaries multiple times (\autoref{S:galactic-realisation}). To avoid clutter, we plot resolvable binaries in the scatter plots only once and ignore the repeated draws. However, we create the histograms taking into account the repeated draws. 

The difference between the \rapid\ and \delayed\ models manifests most prominently for BHs with $3\leq\mbh/\msun\leq5$. 
In the \rapid\ (\delayed) model, about $2.4\%$ ($65\%$) of the resolvable BH-LCs contain BHs in this mass range. If we discard the systems that have undergone CE involving a HG donor, the contrast is even starker. For resolved BH-LCs with BHs in this mass range, while in the \rapid\ model we do not find any BH-MS binaries, in the \delayed\ model BH-MS and BH-PMS contribute almost equally. Note that, even in the \delayed\ model AIC BHs contribute only at about $13\%$ within this $\mbh$ range and the rest come from core-collapse SN. We already highlighted similar trends in \autoref{fig:fixedpop-mbh} in the context of the intrinsic population of BH-LCs, however, it is interesting that this difference is prominent in the resolvable population as well. Detection of BHs with $3\leq\mbh/\msun\leq5$ with both MS and PMS companions in detached configuration would clearly indicate that the so-called mass-gap from core-collapse SN may not be real. Indeed, the recent discovery of BH-LC candidates with BH masses in this range \citep{Thompson2019, Jayasinghe2021} and microlensing analyses \citep{Wyrzykowski2020} provide observational evidence against the existence of a mass gap. 

\begin{figure}
   \plotone{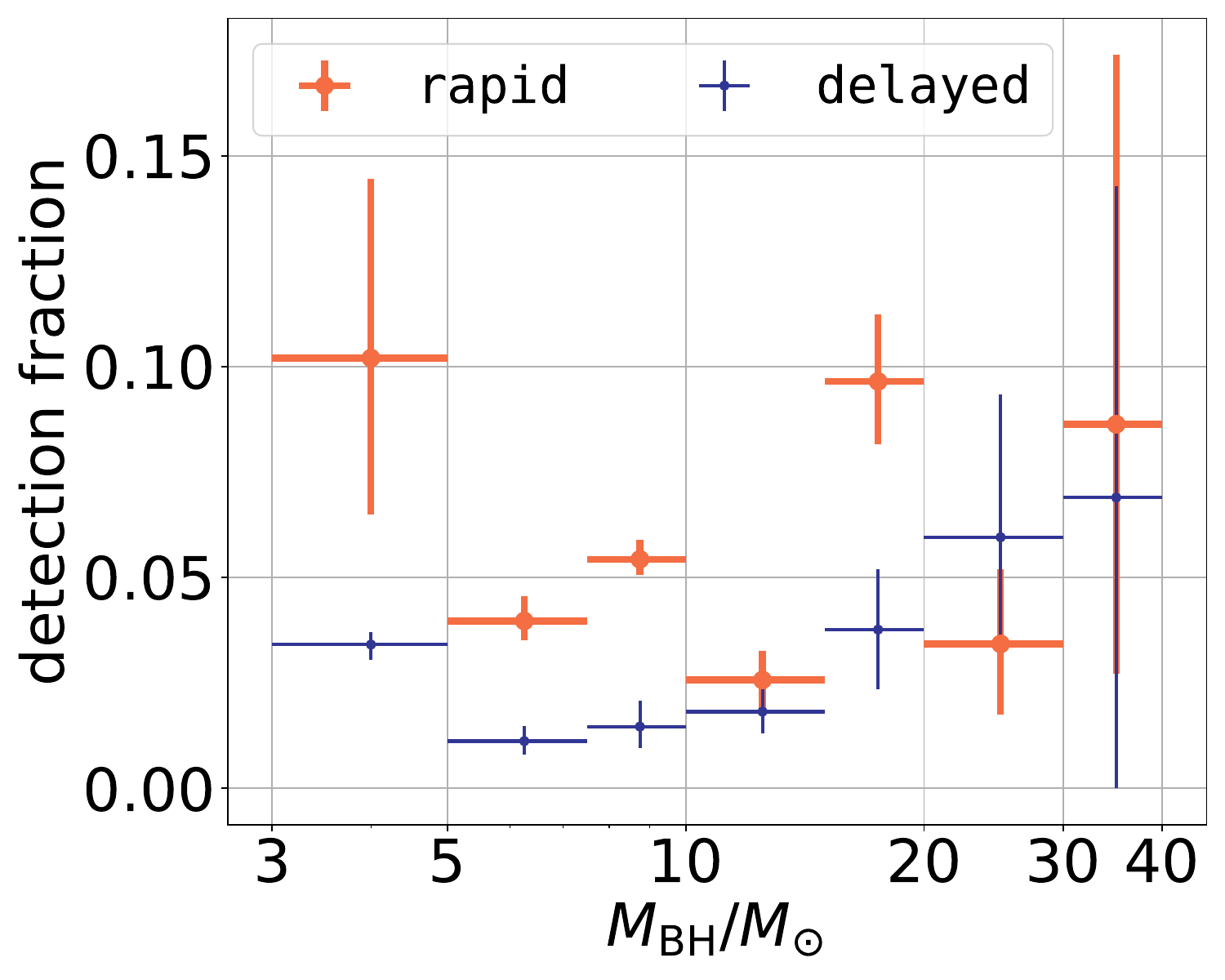}
    \caption{The ratio between the \gaia-resolvable and the intrinsic numbers of detached BH-LC binaries with $\porb/\yr\leq10$ in the Milky Way as a function of $\mbh$. Orange and blue denote \rapid\ and \delayed\ models. The horizontal and vertical errorbars denote the $\mbh$ bin span for the calculation and range between the 10th and 90th percentiles, respectively. We do not find a strong correlation between \gaia's ability to resolve a BH-LC binary and $\mbh$. }
    \label{fig:detection_fraction}
 \end{figure}
 It is interesting that the shapes of the $\mbh$ distributions in our models do not change significantly between the intrinsic and resolvable populations of BH-LC binaries in the Milky Way independent of the adopted SN prescription. This indicates that the ability for \gaia\ to resolve a BH-LC binary do not strongly depend on $\mbh$. Indeed, we find that the ratio between the total numbers of BH-LC binaries in the Milky Way and those that are resolvable by \gaia, do not strongly correlate with $\mbh$ (\autoref{fig:detection_fraction}). This is because \gaia's astrometric selection biases do not depend directly on the BH, rather they depend on the LC and orbital properties, such as $G$ and $\porb$. As a result, the population of BHs resolvable by \gaia\ is expected to exhibit $\mbh$-distribution that is very close to the intrinsic one in the Milky Way. 

The other potentially observable difference between the BH-LCs in the \rapid\ and \delayed\ models stems from the differences in the fallback-dependent natal kicks in the two respective SN mechanisms. The majority of the resolvable systems- about $90\%$ ($95\%$) in the \rapid\ (\delayed) model- go through dissipative binary evolution phases such as CE evolution and RLOF. The final eccentricities are almost entirely dependent on the natal kicks the BHs receive modulated later by tidal cirularisation or RLOF. Because of this, in both the \rapid\ and \delayed\ models, there are prominent peaks near circular orbits. We find that about $87\%$ ($33\%$) of the resolvable BH-LC binaries have $\ecc\leq0.1$ in the \rapid\ (\delayed) model. Similar to the intrinsic population, the resolvable BH-MS binaries in the \delayed\ model show a higher fraction of eccentric systems compared to those in the \rapid\ model. We find that the resolvable systems with $\ecc\geq0.5$ in both \rapid\ and \delayed\ models form predominantly from progenitor binaries that never underwent a CE evolution. Since, this group also consists of relatively wider orbits, they are also less affected by tides after BH formation. In both models, the fraction of eccentric BH-MS binaries is higher than that of eccentric BH-PMS binaries in the resolvable population since the BH-PMS binaries are more strongly affected by tides. Thus future observations of the eccentricity distribution of BH-LC binaries can put constraints on the natal kick strengths for stellar-mass BHs. Such constraints would be complimentary to those based on the locations of BH XRBs away from the Galactic plane \citep[e.g.,][]{Repetto2012,Repetto2015} or those coming from detailed modeling of individual observed BH XRBs \citep[e.g.,][]{Brandt1995,Gualandris2005,Fragos2009,Wong2014}. 
 
Similar to the intrinsic population in the Milky Way, the resolvable population also shows a wide range in metallicities. We also find an expected anti-correlation between metallicity and $\mbh$ since metallicity-dependent stellar winds strongly reduce the maximum BH mass for near solar metallicities \citep[e.g.,][]{Belczynski2010}. Several other expected correlations are apprent in \autoref{fig:resolvable-corner-rapid} and \ref{fig:resolvable-corner-delayed}. For example, longer-period binaries are detected at further distances; all resolvable detached BH-LC binaries reside right of a line representing the limiting angular size of the orbit. Similarly, brighter systems are detected at higher distances. In contrast, no clear correlation is found between $\mbh$ and the distance to the resolvable binary. Some other relations, for example between distance and metallicity, simply come from the fact that we have used initial conditions such that the complex dependence between age-metallicity-distance in the Milky Way is preserved (\autoref{S:initialising}).    

\subsection{Effects of Reddening}
\label{S:MW-reddening}
\begin{figure}
    \plotone{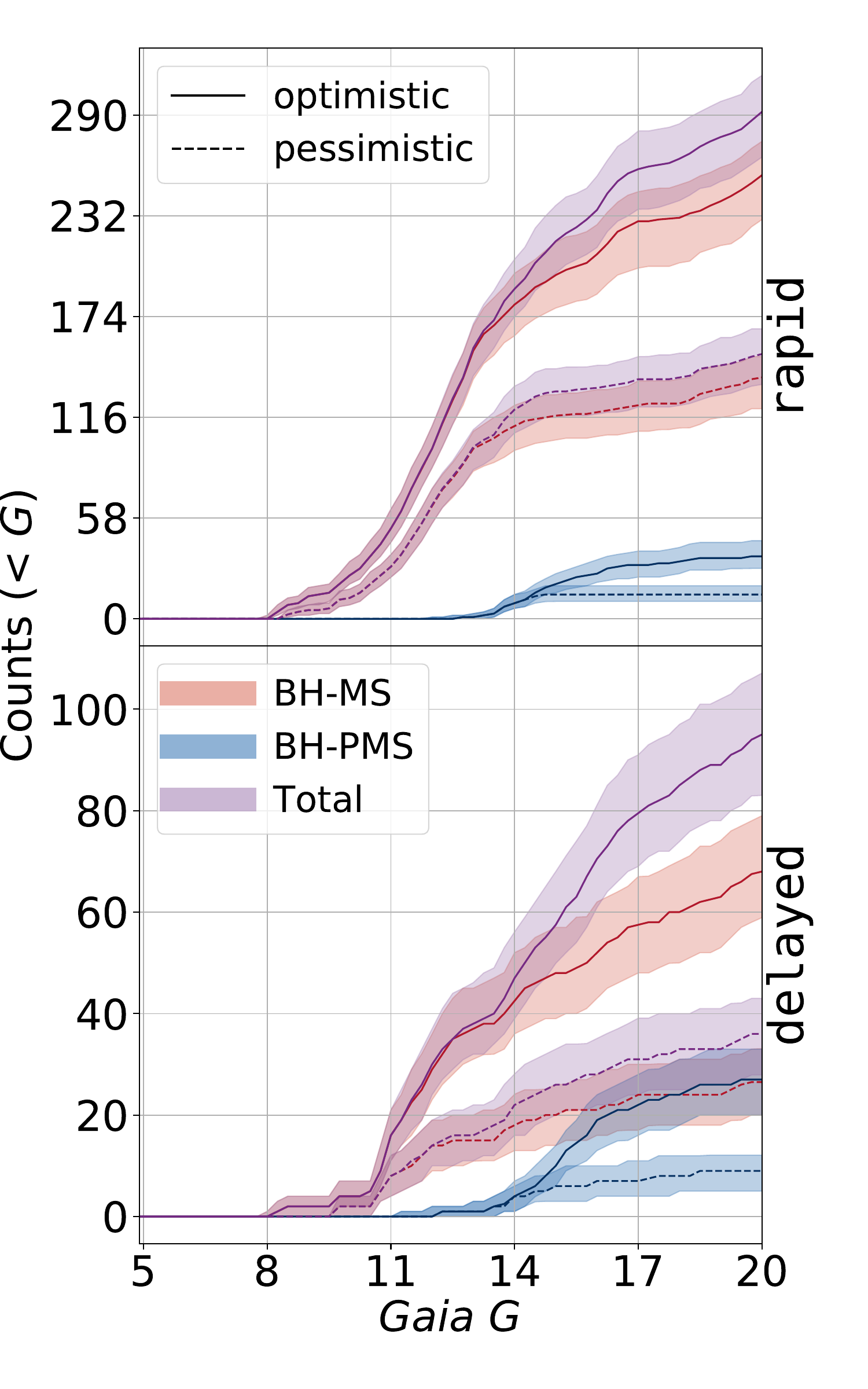}
    \caption{Cumulative number of \gaia-resolvable BH-LC binaries after incorporating the effects of extinction and reddening. The line-styles, colors, and the shades have the same meaning as in \autoref{fig:resolvable-numbers}. After correcting for extinction and reddening, the expected number of detections is between $\approx36$ (\delayed\ model with pessimistic cut) $292$ (\rapid\ model with optimistic cut). We find a higher expected yield for the \rapid\ model compared to the \delayed\ model by a factor of about $3.0$ ($4.2$) using the optimistic (pessimistic) cut. This difference can be attributed to typically lower-mass BHs and larger kicks in case of \delayed. 
    }
    \label{fig:resolvable-numbers-reddening}
\end{figure}
In this section we evaluate the effects of reddening on our estimated yield of detached BH-LC binaries by \gaia. The details of our method to determine extinction and reddening ar described in \autoref{S:reddening}. \autoref{fig:resolvable-numbers-reddening} shows the expected number of detections as a function of \gaia's $G$ magnitude after taking into account the effects of extinction and reddening. We find that reddening and extinction significantly decreases the expected number of detections (\autoref{tab:detection}). After correcting for reddening and extinction we estimate that \gaia\ would resolve between $292^{+21}_{-26}$ (optimistic) and $152^{+14}_{-17}$ (pessimistic) BH-LCs in the \rapid\ model. Whereas, the corresponding numbers for the \delayed\ model are $95^{+12}_{-12}$ (optimistic) and $36^{+7}_{-8}$ (pessimistic). Our extinction- and reddening-corrected estimate for the BH-LC binaries resolvable by \gaia\ provide the most realistic estimates for detection yields. Thus, in the following sections we proceed with the extinction- and reddening-modulated resolvable systems.  

\subsection{Possible astrometric constraints on $\mbh$}
\label{S:MW-conf-BH}
\begin{figure}
    \plotone{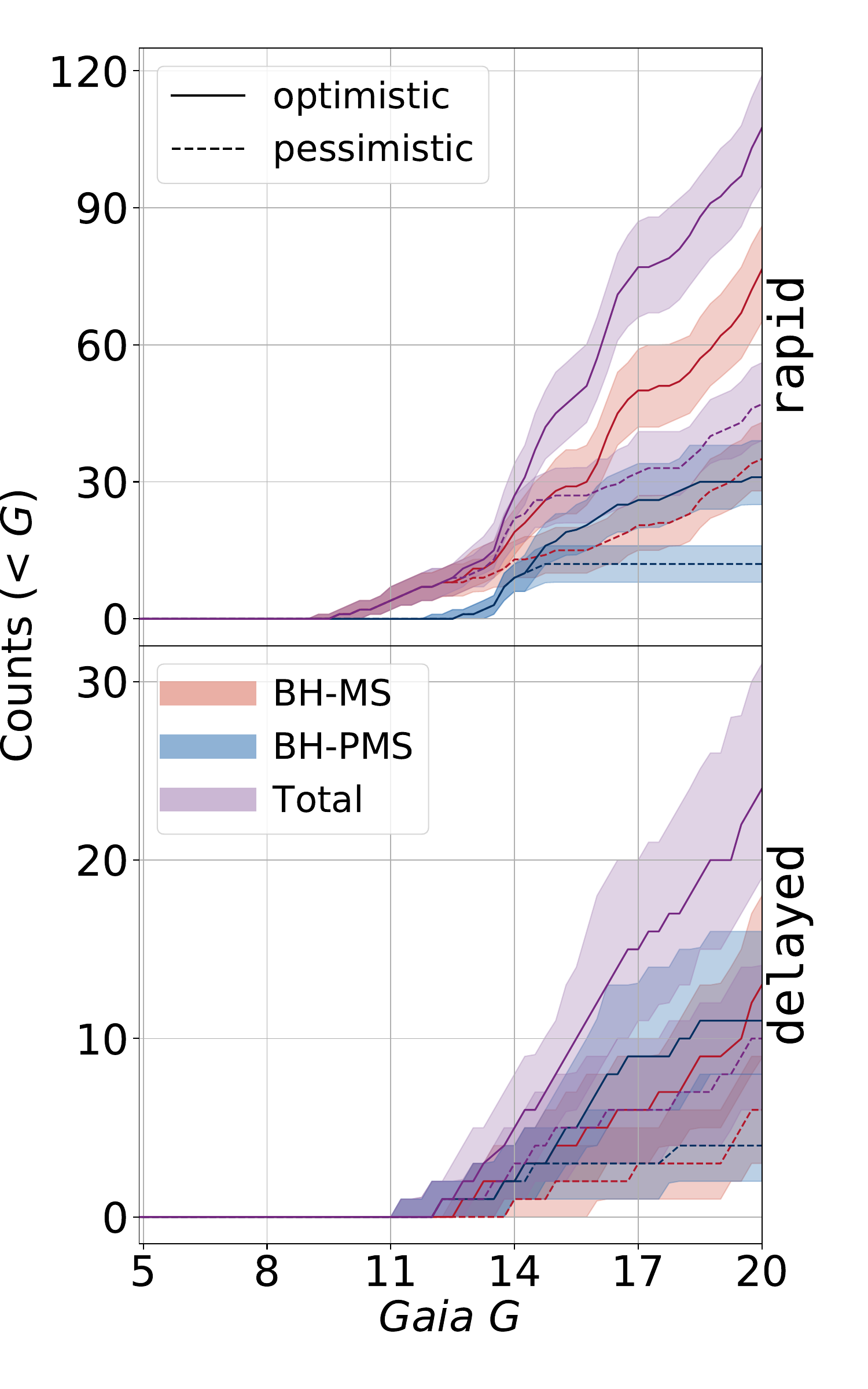}
    \caption{Same as \autoref{fig:resolvable-numbers-reddening}, but after imposing the additional constraints, $\mbh-\Delta\mbh \geq 3\,\msun$ and $\mbh\geq\mlc$ on the extinction- and reddening-corrected detection numbers. We call this subset BH candidates. We estimate $\Delta\mbh$ from \autoref{eq:delta-mbh}. We assume, on an average, $10\%$ error in estimating $\mlc$. These numbers indicate the number of cases where \gaia's astrometry of the LC alone can identify that the dark companion must be a BH.
    }
    \label{fig:resolvable-numbers-confirmed}
\end{figure}
In this section we investigate whether \gaia's astrometry alone can clearly identify the nature of an unseen dark companion to an observed LC as a BH via mass constraints with sufficient accuracy. Note that, in our analysis, the LC exclusively dominates the emission observed by \gaia. Hence, a higher-mass unseen object is most likely to be a dark remnant. This essentially means that if \gaia's astrometry can provide mass constraints that clearly indicate that an unseen companion is more massive than any NSs or the LC mass, it must be a BH candidate. 

By \gaia's final data release, the parallax to identified binaries will be supplied. As a result, the absolute luminosity and hence the mass of the LC can be constrained from the \gaia\ magnitudes using standard stellar evolution models \citep[e.g.,][]{Anders2019,Howes2019}. The constrained mass of the LC and the determined $\porb$ from \gaia\ astrometry can then provide constraints for the unseen dark remnant in the binary. In \citet{Andrews_2019} we showed that this is possible and the accuracy in the mass measurement for the dark component depends on the accuracy in the mass measurement of the LC and the accuracy of the along and across scans for the source. The analytic expression and the details of the calculation for mass errors of the BHs are given in \autoref{S:mass-constraint}. 

In order to find the subset of BH-LC binaries where \gaia's astrometry alone can reveal that the dark remnant is a BH, in addition to all of \gaia's detectability considerations (\autoref{S:resolve}) and effects of extinction and reddening (\autoref{S:reddening}), we employ two additional filters, (i) $\mbh>\mlc$ and (ii) $\mbh - \Delta \mbh \geq 3\,\msun$. The former condition is to rule out confusion potentially created by very high-mass bright primary in orbit with another less bright secondary of mass above $3\,\msun$ in observed systems. We denote this subset as BH candidates. Note that our constraint of $\mbh-\Delta\mbh\geq3\,\msun$ is conservative and consistent with the assumed boundary between NSs and BHs in our simulations. We do not suggest that NSs can be as massive. The estimated masses of the heavy NSs to date are lower than $3\,\msun$ \citep[e.g.,][]{Cromartie2020,Biswas2021}.

\begin{figure}
    \plotone{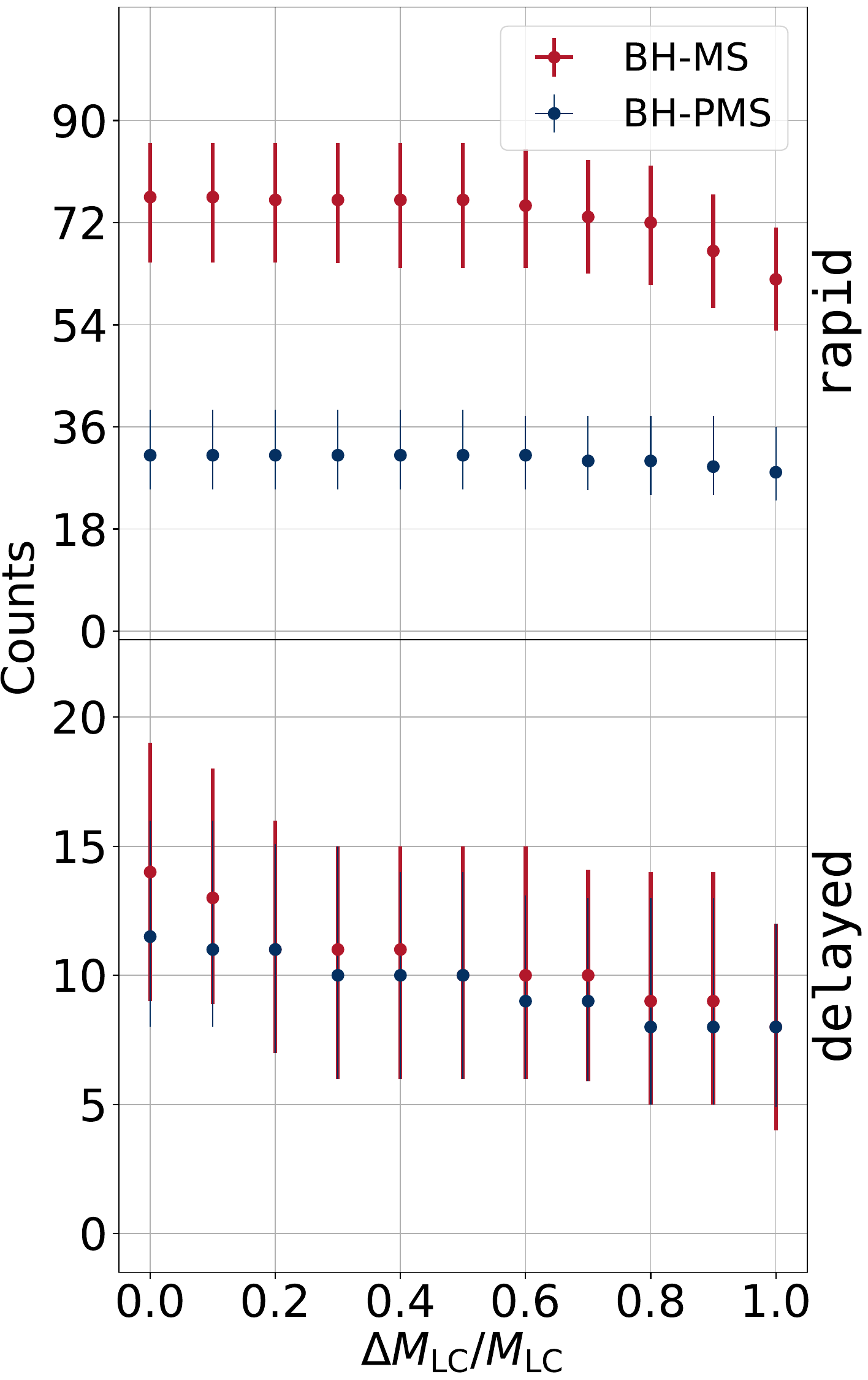}
    \caption{Number of BH candidates as a function of the adopted value of $\Delta\mlc/\mlc$. Red (blue) dots and errorbars denote the median, and the range between the 10th and 90th percentiles for BH-MS (BH-PMS) binaries. The number of BH candidates remains within the levels of statistical fluctuations throughout the full range in $\Delta\mlc/\mlc$.
    }
    \label{fig:resolvable-numbers-confirmed-del-mlc}
\end{figure}

\autoref{fig:resolvable-numbers-confirmed} shows the expected number of BH-LC binaries as a function of \gaia's $G$ magnitude, where astrometry alone can determine that the mass of the unseen primary to a secondary LC is $\geq 3\,\msun$. We find that all of the above stringent criteria are satisfied for $\approx37\%$ ($25\%$) of reddening-corrected resolvable systems in the \rapid\ (\delayed) model. 
According to our optimistic threshold for \gaia's astrometric precision, the total numbers of resolvable BH-LC binaries with astrometric mass measurements strongly indicating a BH as the dark object, is $107^{+11}_{-12}$ ($24^{+7}_{-5}$) for the \rapid\ (\delayed) model. Based on the pessimistic threshold, the corresponding number is $47^{+9}_{-8}$ ($10^{+4}_{-4}$) (\autoref{tab:detection}). 

While the results shown in \autoref{fig:resolvable-numbers-confirmed} uses the fiducial value $\Delta\mlc/\mlc = 0.1$, we note that the precision in mass estimates of isolated single or binary stars can vary widely depending on the system, data quality, as well as the measurement technique \citep[e.g.,][]{Torres2002,Gallenne2016,Pavlovski2018,Brogaard2018,Rendle2019}. Our \autoref{eq:delta-mbh} provides a straightforward way to use any desired value of $\Delta\mlc/\mlc$ to ultimately update the estimated $\Delta\mbh/\mbh$ through \gaia's astrometry. Nevertheless, by varying $\Delta\mlc/\mlc$ between the full range of zero and one, we find that the estimated number of BH candidates do not depend strongly on the chosen value of $\Delta\mlc/\mlc$ (\autoref{fig:resolvable-numbers-confirmed-del-mlc}). In fact, throughout the full range, the estimated numbers stay within the range due to statistical fluctuations for any one of the chosen values of $\Delta\mlc/\mlc$ including our fiducial value of 0.1. The lack of sensitivity towards the adopted value of $\Delta\mlc/\mlc$ can be easily understood from \autoref{eq:delta-mbh}. To identify BH candidates, we impose the condition $\mbh\geq \mlc$. This makes the term involving $\Delta\mbh/\mbh$ less significant compared to the other term involving astrometric precision.

\section{Promise from followup measurements beyond astrometry}
\label{S:beyond}
While our primary focus of this work is on the ability of \gaia\ to detect and characterise detached BH-LC binaries using astrometry alone, of course, once identified by \gaia's astrometry, followup observations of these binaries can significantly improve the characterisation of the binary properties. In this section we highlight a couple of these promising avenues. 

\subsection{Radial velocities using \gaia's spectroscopy}
\label{S:MW-RV}
The most readily available avenue to improve characterisation of astrometrically identified candidates of interest is RV measurements by \gaia\ itself. \gaia\ contains a RV spectrometer which provides spectral resolution $R\sim5000$ (low) and $R\sim11500$ (high) for stars with $G\leq17$ and $G\leq11$, respectively. We find that \gaia's high-$R$ spectra may not be very useful in characterising BH-LC binaries. While the high-$R$ allows measurement of lower RV semiamplitudes ($K$), the stronger constraint on $G$ prohibitively limits the number of available BH-LC binaries in the Milky Way.   

\begin{figure}[h]
    \centering
    \plotone{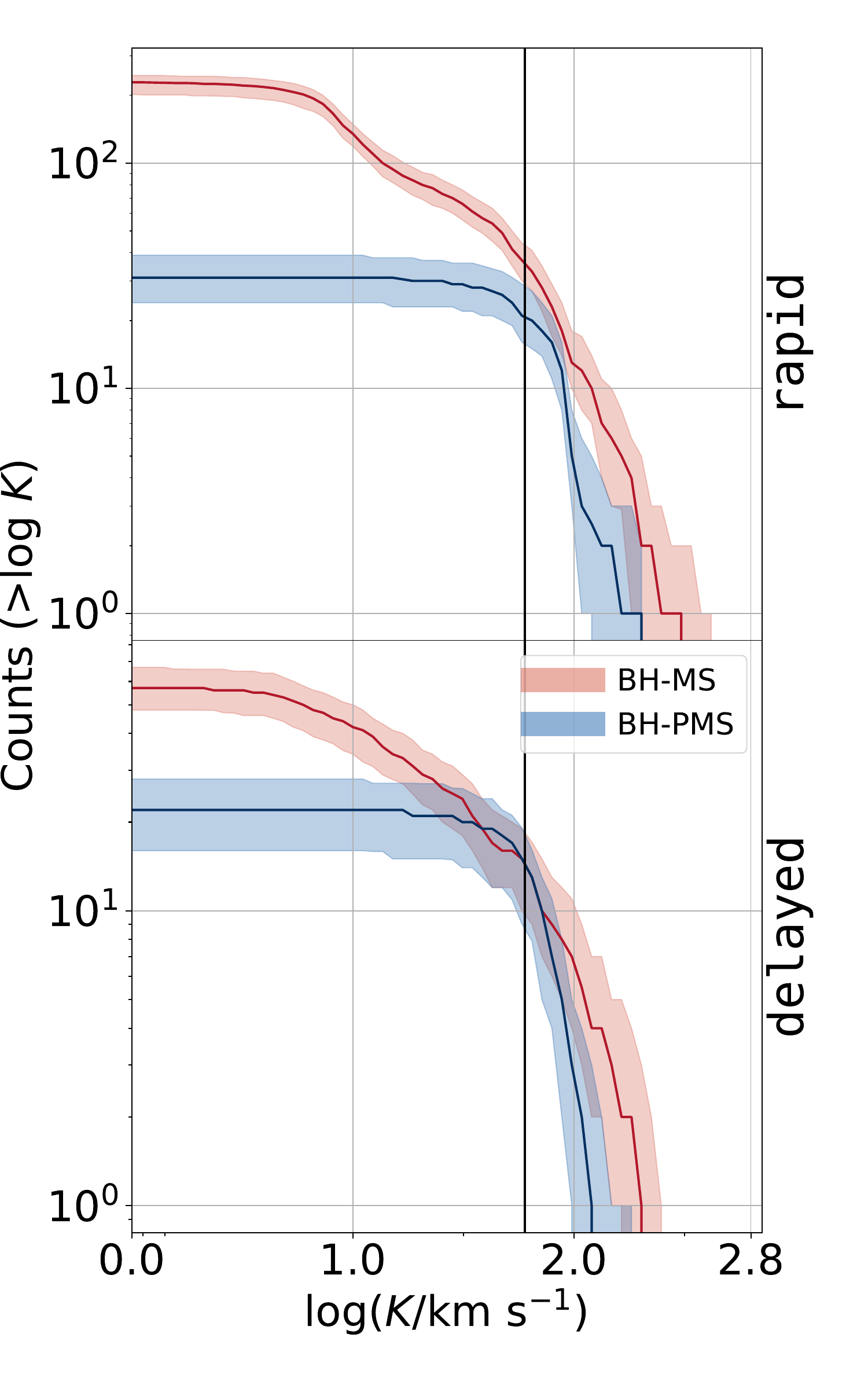}
    \caption{Reverse cumulative distribution of the RV semi-amplitude ($K$) for detached BH-LC binaries resolvable by \gaia\ (using the optimistic threshold for astrometric resolution) with $G<17$ after correcting for extinction and reddening. The vertical line denotes the level of $K$ resolved by \gaia's low spectral resolution ($R$) data. Red and blue denote BH binaries with MS and PMS companions. Lines and shaded regions around the lines denote the median and the spread between the 10th and 90th percentiles. 
    }
    \label{fig:resolvable-rv}
\end{figure}
\autoref{fig:resolvable-rv} shows the distribution RV semiamplitude ($K$) for the reddening- and extinction-corrected, resolvable BH-LC binaries with $G<17$. We find that in the \rapid\ model $\sim35$ BH-MS and $21$ BH-PMS binaries could have high enough $K$ to be resolved by \gaia's low-$R$ spectroscopy (\autoref{fig:resolvable-rv}). In comparison, in the \delayed\ model $\sim14$ BH-MS and $\sim15$ BH-PMS binaries are expected to have high enough $K$ to be resolved by \gaia's low-$R$ spectra. Using \gaia's high-$R$ spectra, these numbers are at most a few in both \rapid\ and \delayed\ models. As can be seen in \autoref{fig:resolvable-rv}, $K$ for \gaia-resolvable BH-LC binaries is rather large. Hence, a compromise in the spectral resolution matters less than the loss of available systems in the Milky Way from a stronger constraint in $G$. 

Overall, we find that a little over $35\%$ of the detached \gaia-resolvable (reddening-corrected) BH-LC binaries will have some constraints on $\mbh$ either through astrometry (as described in \autoref{S:MW-conf-BH}) or through RV measurements using \gaia's low-$R$ spectra. Although, we limit our discussion to RV measurements using \gaia's spectrometer, of course, once the BH binary candidates are identified, a higher fraction of systems may be constrained via RV followup using other higher-$R$ telescopes that can probe fainter stars. We also caution the readers that in our quick analysis we did not consider the details that affect the feasibility of RV measurements of a particular LC which include availability of lines to observe within the band of the spectrometer, Galactic position, and stellar activity. 

\begin{figure}[h]
    \centering
    \plotone{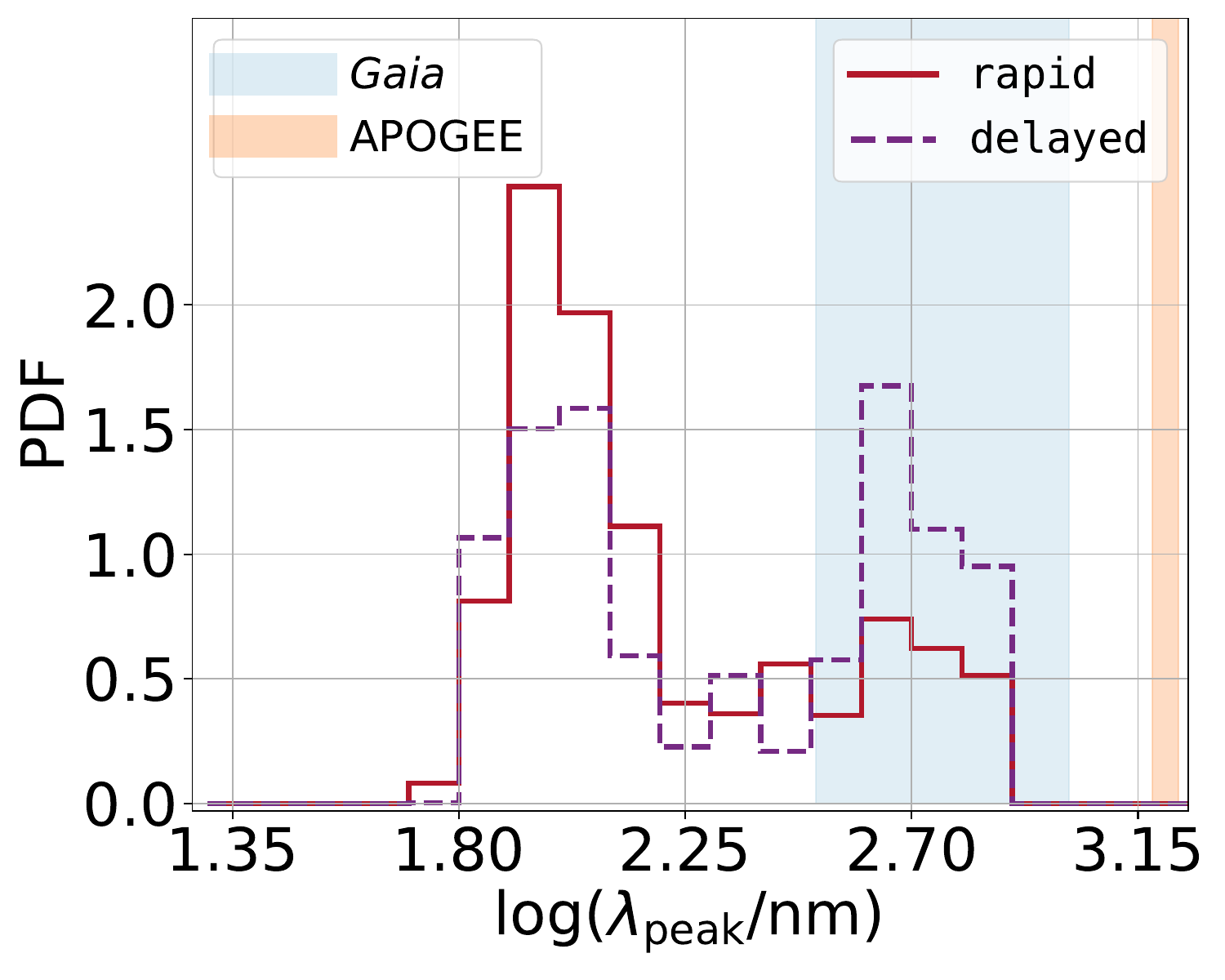}
    \caption{Distribution of the wavelengths ($\lambda_{\rm{peak}}$) corresponding to the peak in the blackbody spectral-energy distribution for the resolvable \rapid\ (red) and \delayed\ (purple) binaries.  Blue and orange vertical bands show the ranges of bandpass filters for \gaia\ ($330$--$1050\,\rm{nm}$) and APOGEE ($1.51$--$1.70\,\micron$). The wavelength band for \gaia\ is more favorable for the majority of the BH-LC binaries compared to that of APOGEE. 
    }
    \label{fig:resolvable-lambda-peak}
\end{figure}
Till date the discovered detached BH-LC binaries have used data from multi-epoch spectroscopic surveys by the MUSE \citep[e.g.,][]{Giesers2018,Giesers2019} and APOGEE collaborations \citep[e.g.,][]{Thompson2019,Jayasinghe2021}. MUSE specifically focuses in crowded fields such as globular clusters, whereas, primary APOGEE targets are in the field. Hence, it is interesting to compare \gaia's capabilities with that of APOGEE in the context of detecting BH-LC binaries using RV measurements. In order to find the number of \gaia-resolvable BH-LC binaries APOGEE may also detect, we take the subset of reddening-corrected \gaia-resolvable BH-LC binaries in our models and apply two additional filters relevant for APOGEE. In particular, we use $J-K_s \leq 0.5$ and $12.2 \leq H \leq 14$ \citep{Zasowski2013,Price-Whelan2020}. We find that APOGEE cannot detect any BH-MS binaries in our \gaia-resolvable populations and may detect $\sim1$ ($\sim3$) BH-PMS binaries in the \rapid\ (\delayed) model.
Note that, both APOGEE detections of BH candidates so far contain PMS companions \citep[e.g.,][]{Thompson2019,Jayasinghe2021}. The reason why \gaia\ is expected to be more prolific in discovering BH-LC candidates, especially, BH-MS binaries, can be understood by comparing the filters of \gaia\ and APOGEE (\autoref{fig:resolvable-lambda-peak}). APOGEE is sensitive to much cooler stars compared to \gaia. Most of the detached BH-LC binaries that \gaia\ can resolve are too faint in APOGEE. 

\subsection{X-ray counterparts}
\label{S:MW-X-ray}
\begin{figure}[h]
    \centering
    \plotone{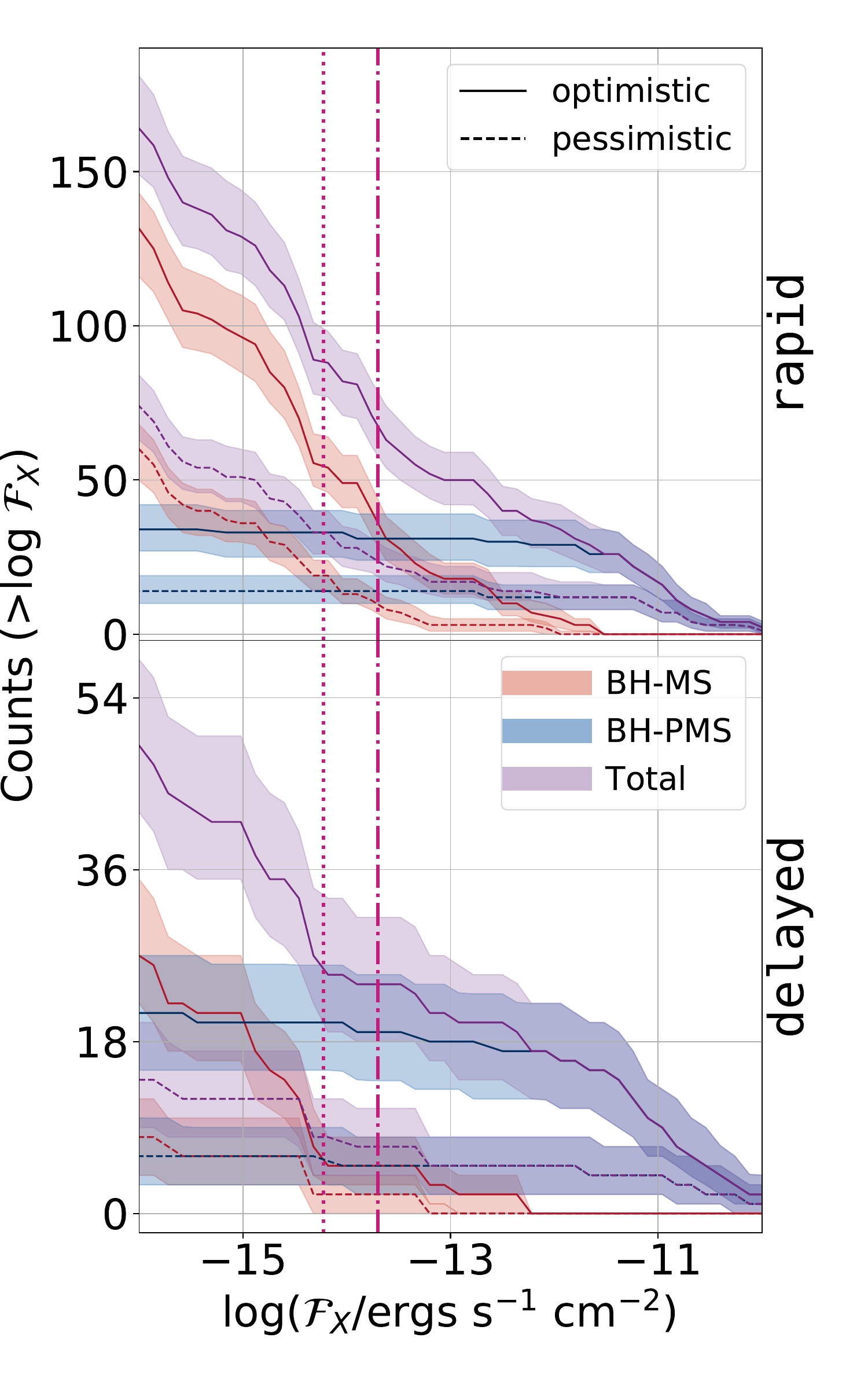}
    \caption{Reverse cumulative distributions of reddening-corrected,  resolvable, detached BH-LC binaries as a function of the extinction and reddening corrected X-ray flux $\mathcal{F}_{\rm{x}}$ in the band $0.1-10 {\rm{keV}}$. All systems shown here are wind-fed systems. The vertical pink lines denote $\mathcal{F}_X=6\times10^{-15}\,\ergspers\ {\rm{cm^{-2}}}$ (dotted) and $\mathcal{F}_X=2\times10^{-14}\,\ergspers\ {\rm{cm^{-2}}}$ (dash-dot), our adopted lower limits for detectability by Chandra and eROSITA, respectively. Red, blue, and purple denote BH-MS, BH-PMS, and all BH-LC binaries. Top and bottom panels are for the \rapid\ and \delayed\ models.
    }
    \label{fig:resolvable-X-ray}
\end{figure}
A fraction of the reddening-corrected, resolvable,  detached BH-LC binaries may have X-ray counterparts if the stellar wind from the LC is accreted by the BH at a sufficiently high rate. 
We can estimate the bolometric X-ray luminosity for these wind-fed systems as-
\begin{equation}
    \label{eq:lumin_x_ray}
    L_{X} = \varepsilon\frac{G \mbh\dot{M}_{\rm{acc}}}{R_{\rm{acc}}},
\end{equation}
where $\dot{M}_{\rm{acc}}$ is the accretion rate of the BH, $\racc$ is the accretion radius, and $\varepsilon$ is an efficiency parameter for the conversion of gravitational binding energy to radiation. For our calculations we assume, $\racc\approx3\times R_{\rm{sc}}$ ($R_{\rm{sc}}$ is the Schwarzschild radius) and $\dot{M}_{\rm{acc}}\sim 10^{-12}$ -- $10^{-8}\,\msun\ \yr^{-1}$. Since $\dot{M}_{\rm{acc}}$
for the resolvable binaries is expected to be low, $\leq 10\%$ of the Eddington rate, we adopt that the accretion process is in the regime of the advection dominated accretion flow (ADAF). We adopt the prescription given in \citet[][their equation 11]{Xie2012} to calculate the efficiency parameter. In particular, assuming the viscosity parameter $\alpha=0.1$, viscous heating parameter $\delta=0.5$, we find $\varepsilon$ ranges from $10^{-5}-0.082$ based on the ratio of $\dot{M}_{\rm{acc}}$ to the Eddington mass accretion rate. 

We assume that the bolometric X-ray luminosity calculated using \autoref{eq:lumin_x_ray} includes X-rays in the energy range $0.1-500{\rm{keV}}$. Since X-ray observatories like Chandra and eRosita are sensitive in the energy band $0.1-10 {\rm{keV}}$, we evaluate the incident flux $\mathcal{F}_{\rm{X}}$ in the energy band $0.1-10 {\rm{keV}}$ assuming a power law spectrum for X-rays,  
\begin{equation}
    N(E) \propto E^{-\Gamma}, 
\end{equation}
with photon index $\Gamma=2$ \citep[][]{Yang2015}. 
Furthermore, we correct $\mathcal{F}_{\rm{X}}$ for interstellar extinction and reddening using  
\begin{equation}
    \mathcal{F}_{\rm{X}} = \int_{0.1}^{10} e^{\sigma_{\rm{ISM}}(E)N_{H}}EN(E) \,dE 
\end{equation}
where $\sigma_{\rm{ISM}}(E)$ is the total photoionization cross section of ISM taken from \citet[][]{Wilms_2000}, and $N_{\rm{H}}$ is the hydrogen column density between the source and the observer. We estimate $N_{\rm{H}}$ simply from the position-dependent $A_v$ obtained using \dustmaps: 
\begin{equation}
    \frac{N_{H}}{A_v}=2.2\times10^{21}{\rm{cm}}^{-2}
\end{equation}
\citep[][]{Guver2009}.

We show the reverse cumulative numbers of reddening-corrected \gaia-resolvable detached BH-LC binaries as a function of $\mathcal{F}_X$ in \autoref{fig:resolvable-X-ray}. We adopt, the detection limits of Chandra and eROSITA as $\mathcal{F}_X\geq6\times10^{-15}\,\ergspers\ {\rm{cm^{-2}}}$ \citep[based on CSC2.0 catalog][]{Weisskopf2000} and $\mathcal{F}_X\geq2\times10^{-14}\,\ergspers\ {\rm{cm^{-2}}}$ \citep[][]{Merloni2012}, respectively. 
We find that the number of reddening-corrected, \gaia-resolvable, detached BH-LC binaries expected to have X-ray counterparts in eROSITA are $68^{+9}_{-10}$ ($24^{+6}_{-5}$) and $24^{+7}_{-6}$ ($7^{+4}_{-3}$) for the \rapid\ and \delayed\ models adopting the optimistic (pessimistic) astrometric threshold. The corresponding numbers for Chandra are $88^{+11}_{-11}$ ($33^{+7}_{-7}$) and $25^{+8}_{-6}$ ($8^{+4}_{-4}$), respectively. Interestingly, although further characterisation is needed to ascertain the nature of these sources several studies have reported possible BH-LC candidates by cross-matching \gaia\ astrometry and X-ray counterparts \citep[e.g.,][]{Gandhi2020,Price-Whelan2020}. 

\section{Summary and Discussion}
\label{S:discussion}
In this paper we have investigated the possibility for \gaia\ to detect a population of detached BH-LC binaries using astrometry. Using a state-of-the-art population synthesis code \cosmic\ \citep{Breivik2020} and realistic stellar ages, metallicities, and Galactic positions adopted from the simulated Milky Way-mass galaxy \textbf{m12i} from the Latte suite of FIRE-2 simulations \citep{Wetzel2016, Hopkins2018, Sanderson2020}, which take into account the observed complex correlations between these parameters, we have created highly realistic present-day populations of BH-LC binaries expected to be found in the Milky Way adopting two widely used SN prescriptions, rapid and delayed \citep[e.g.,][; \autoref{S:methods}]{Fryer2012}. 

We have presented several relevant properties of our simulated present day populations of BH-LC binaries in the Milky Way, including the component masses, orbital properties, age, and metallicity (\autoref{S:fixedpop}). We summarize these findings below.

\begin{itemize}
    \item We find that, intrinsically, the Milky Way is host to $9,184^{+120}_{-133}$ ($7,091^{+113}_{-100}$) detached BH-LC binaries at present with $\porb/\yr\leq10$ based on the \rapid\ (\delayed) model (\autoref{tab:detection}). 
    \item The orbital period distribution, regardless of SN prescription, is bimodal with peaks corresponding to BH-LC populations with progenitors which have undergone at least one CE phase and those which have not experienced a CE. The extended \gaia\ mission lifetime of $10\,\yr$ allows for characterization of both populations.
    \item The BH mass distribution depends strongly on the SN prescription choice. The rapid prescription produces BH populations with a mass gap between $3$--$5\msun$, while the delayed prescription does not.
    \item BHs with masses $\lesssim10\,\msun$ have orbital periods $<10\,\yr$ because of the correlation between natal kick strength and BH mass which unbinds BH-LC progenitors in wider orbits. If \gaia\ discovers BH-LCs with BH masses outside of this range, constraints on the correlation between natal kick strength and BH mass can be imposed.
    \item Eccentricity is a strong tracer of both natal kick strength and tidal circularisation. The majority of BH-PMS systems are circularized through tides, while $\sim1\%$ ($7\%$) of BH-MS have $Ecc>0.1$ in the \rapid\ (\delayed) models. 
    \item The age and metallicity of BH-LC binaries in both the \rapid\ and \delayed\ models are broadly distributed, indicating the potential to observationally constrain a metallicity-dependent BH mass distribution.
\end{itemize}

In order to determine the subset of BH-LC population resolvable by \gaia, we create $200$ realisations of the Milky Way sampling from our simulated intrinsic populations of BH-LC binaries for the \rapid\ and \delayed\ models. Each of these realisations preserve the observed correlations between the age, metallicity, and Galactic positions of stars in the Milky Way as well as the expected number and parameter distributions of the BH-LC population (\autoref{S:galactic-realisation}). For each of these realisations and each SN model, we investigate detectability using \gaia's astrometry (\autoref{S:detectability}). We summarise our findings below.

\begin{itemize}
    \item We predict that $\sim180$---$470$ BH-LCs have $G<20$, $\porb < 10\,\yr$, and $\alpha > \sigma_\xi$ depending on the choice of SN prescription. These numbers are reduced to $\sim60$---$240$ if we apply a more pessimistic astrometric threshold $\alpha > 3\sigma_\xi$.
    \item Extinction and reddening from interstellar dust reduces the expected yield of \gaia-resolvable BH-LCs. The expected yield after reddening correction is $292$ ($95$) for the \rapid\ (\delayed) model using the optimistic threshold; for the pessimistic threshold, these numbers are $152$ ($36$).  
    \item We find that \gaia's astrometry alone can identify BH candidates with some certainty for $\approx107$ ($\approx24$) BH-LC binaries in our \rapid\ (\delayed) model. In addition to being resolvable (after reddening correction) by \gaia's astrometry, these systems satisfy 
    $\mbh-\Delta\mbh\geq3\,\msun$ and $\mbh\geq\mlc$. 
    \item Independent of the adopted SN prescription, $\sim50\%$ of the resolvable BH-MS progenitors undergo at least one CE phase while $\gtrsim90\%$ of the resolvable BH-PMS experience at least one CE phase.
    \item The predicted intrinsic vs \gaia-resolvable Milky-Way populations of BH-LCs do not show a strong correlation with BH mass. This suggests that \gaia's observational selection does not strongly bias our view of the BHs in detached binaries in the Galaxy.
    \item Intrinsic correlations in BH mass and eccentricity with SN prescription and BH mass with metallicity persists in the \gaia-resolvable population, highlighting opportunities to provide constraints on uncertain SN physics with future population detections.
\end{itemize}

In this paper we have focused on the promise from \gaia's astrometry. Nevertheless, we point out that the orbital solutions and component characterisation of the BH-LC population potentially resolvable by \gaia\ astrometry can be improved significantly by analysing \gaia's RV data as well as RV followup using other telescopes. 
We find that the RV semi-amplitude $K$ can be several to hundreds of $\kms$ for the reddening-corrected resolvable population of BH-LC binaries in our models (\autoref{S:MW-RV}, \autoref{fig:resolvable-rv}). \gaia's own low-resolution spectroscopy can resolve the RV for $\sim 56$ ($\sim29$) BH-LC binaries within the reddening-corrected resolvable population in our \rapid\ (\delayed) model. 
In both models, about $35\%$ of the extinction- and reddening-corrected \gaia-resolvable BH-LCs either have high enough $K$ to be resolved by \gaia's low-R spectra or they can be identified as BH candidates through astrometric mass constraint independent of the adopted threshold for astrometric precision. 

At present, APOGEE data provides the most relevant all-sky RV survey we can compare with. We find that the APOGEE RV survey may not be as prolific as what is expected of \gaia's RV data based on our models. This is because most resolvable BH-LC binaries, except for some BH-PMS binaries, are expected to be too faint in the APOGEE bands (\autoref{fig:resolvable-lambda-peak}). 

Our models show that between $\sim 7$ to $88$ (depending on the adopted SN prescription and astrometric threshold) of the reddening-corrected \gaia-resolvable detached BH-LC binaries may also have X-ray counterparts as wind-fed systems potentially detectable by Chandra and eROSITA. In addition to these detached wind-fed systems, we estimate that the orbital motion of the PMS companion may be resolvable by \gaia\ astrometry for $\approx 76$  BH-PMS binaries currently transferring mass via ROLF in both \rapid\ and \delayed\ models. We have ignored these binaries in this work to focus on the detached BH-LC binaries. We suspect that finding an orbital solution astrometrically for the non-detached systems will be significantly more challenging compared to doing so for the detached systems even if \gaia\ astrometry can resolve the orbital motion of the LC. However, these systems can be interesting candidates for multi-messenger studies potentially creating a population of BH-LC binaries that connect the populations of BH X-ray binaries and detached BH binaries potentially resolvable in large numbers by \gaia\ soon.  

In conclusion, our models suggest that \gaia\ data could dramatically improve our understanding of the properties and demographics of BH binaries in the Milky Way. Such detections are expected to have few selections biases depending on the BH mass and would provide unprecedented constraints on BH-progenitor properties by constraining the stellar properties of the observed LC.  

\acknowledgements{We thank the referee for constructive comments. CC acknowledges support from TIFR's graduate fellowship. SC acknowledges support from the Department of Atomic Energy, Government of India, under project no.  12-R\&D-TFR-5.02-0200 and RTI 4002. JA acknowledges funding from CIERA and Northwestern University through a postdoctoral fellowship. The Flatiron Institute is supported by the Simons Foundation.}

\section{Code and Data availability}
The data and code used for this study is freely accessible on \texttt{Zenodo} \citet[][]{chawla_chirag_2021}.

\software{\texttt{Astropy}\ \citep{astropy:2013, astropy:2018}; \cosmic\ \citep{Breivik2020}; \dustmaps\ \citep{Bovy2016}; \texttt{isochrones}\ \citep{2015ascl.soft03010M}; \texttt{matplotlib}\ \citep{matplotlib}; \texttt{numpy}\ \citep{numpy}; \texttt{scipy}\ \citep{scipy}}

\bibliographystyle{aasjournal}
\bibliography{main}  

\begin{figure}
	  \plotone{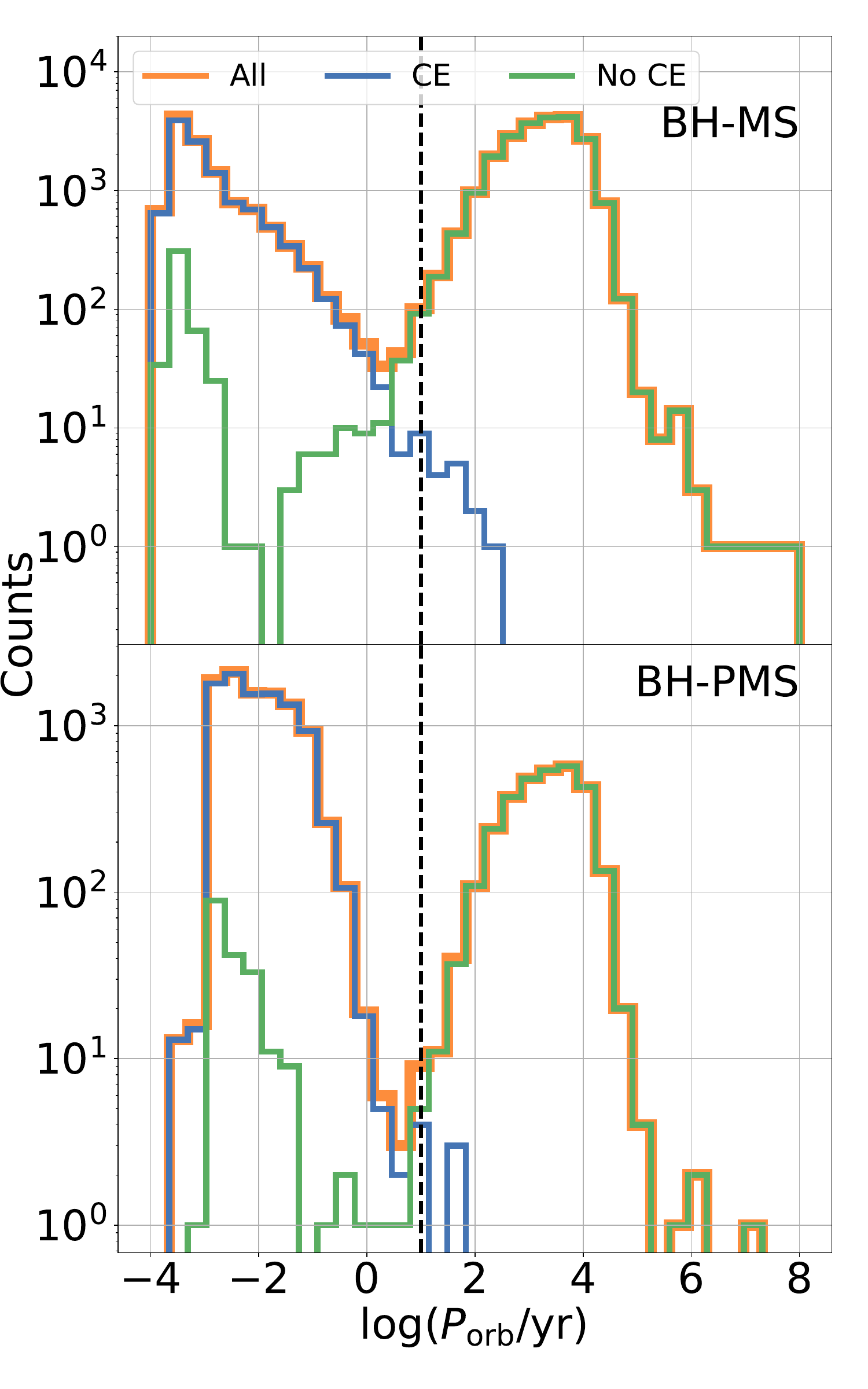}
    \caption{Same as \autoref{fig:fixedpop-porb} but for the BH-LCs in our \delayed\ model. 
    }
    \label{fig:fixedpop-porb-delayed}
\end{figure}

\appendix
Here we present selected results from our \delayed\ model.  \autoref{fig:fixedpop-porb-delayed} shows the $\porb$ distribution for the \delayed\ model for our converged population of BH-LC binaries. The distribution of $\porb$ in the \delayed\ model is very similar to the same for the \rapid\ model (\autoref{fig:fixedpop-porb}). 

\begin{figure}
    \plotone{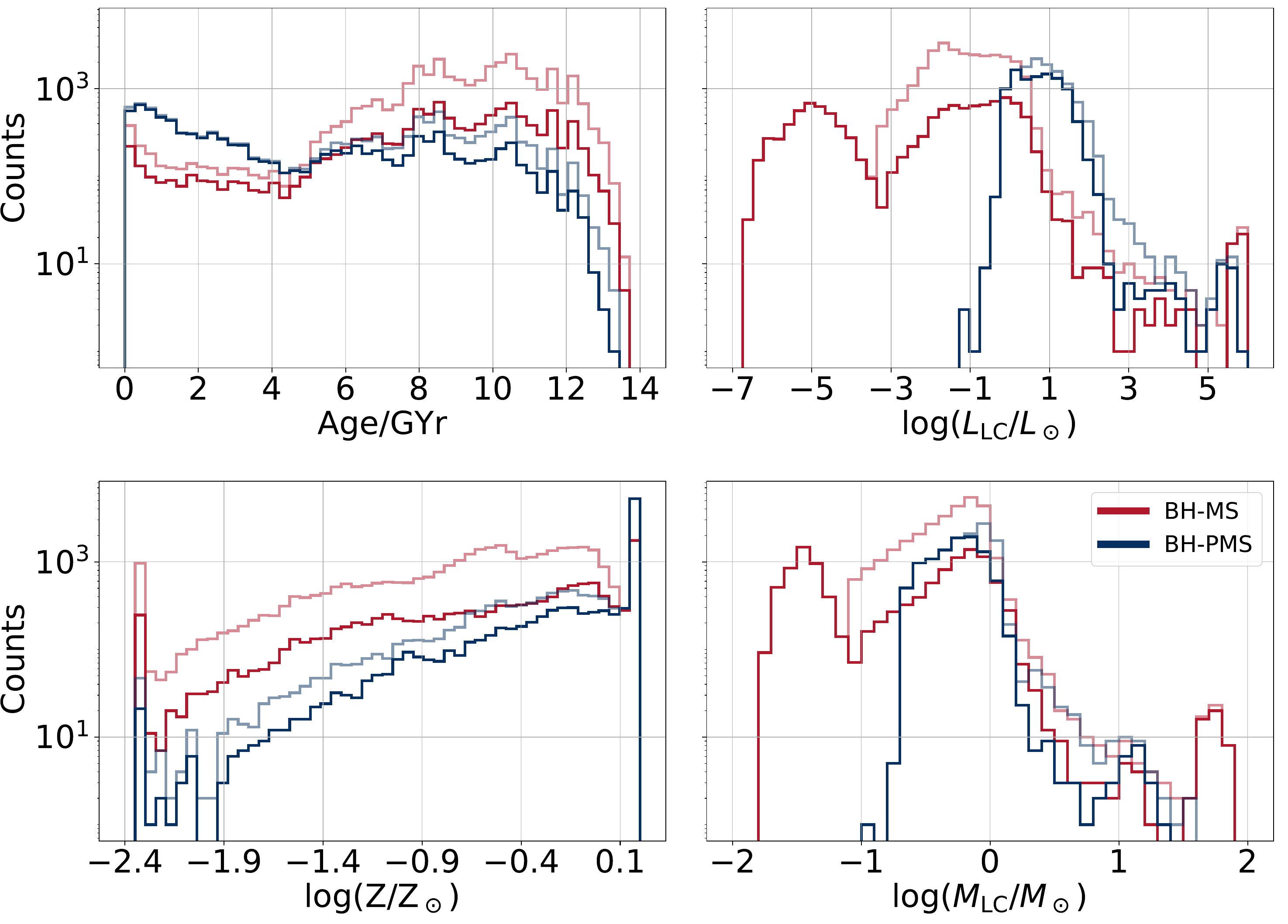}
    \caption{Same as \autoref{fig:fixedpop-otherprop} except for the BH-LC binaries in our \delayed\ model. }
    \label{fig:fixedpop-otherprop-delayed}
\end{figure}

\autoref{fig:fixedpop-otherprop-delayed} shows the relevant LC properties for the intrinsic population of BH-LC binaries in our \delayed\ model. 

We have used propagation of error to estimate $\Delta\mbh$ from the analytic expression for the astrometric mass function for the LC's motion (without the $\sin i$ degeneracy) given in \autoref{eq:mf-fractional}. \citet{Andrews_2019} derived this expression using a suite of Markov-chain Monete Carlo parameter-estimation exercise using injected LC motion as would be seen by \gaia\ assuming a $5$-$\yr$ mission duration. We have updated their expression for a $10$-$\yr$ \gaia\ mission for our purpose. Below we show the steps used to derive \autoref{eq:delta-mbh} from \autoref{eq:mf-fractional}. 
For simplicity, we write the astrometric mass function without the $\sin\ i$ term as- 
\begin{eqnarray}
    \label{eq:appendix1}
    x & \equiv & \frac{\mbh^3}{(\mbh+\mlc)^2} \nonumber\\
    & = & \mbh\left(1+\frac{\mlc}{\mbh}\right)^{-2} \nonumber\\
    & = & \mbh \times y^{-2},
\end{eqnarray}
where, we have used $y = (1+\mlc/\mbh)$. Taking $\ln$ on both sides of \autoref{eq:appendix1} and differentiating we get- 
\begin{equation}
    \label{eq:appendix2}
    \frac{\partial x}{x} = \frac{\partial \mbh}{\mbh} - 2\frac{\partial y}{y}
\end{equation}
Since, $y$ is a function of both $\mbh$ and $\mlc$, we evaluate $\partial y/y$ as-
\begin{eqnarray}
    \label{eq:appendix3}
    \frac{\partial y}{y} & = & \left.\frac{dy}{d\mbh}\right|_{\mlc}\frac{\partial \mbh}{y} + \left.\frac{dy}{d\mlc}\right|_{\mbh}\frac{\partial \mlc}{y} \nonumber\\
    & = & -\frac{2\mlc}{\mbh^2}\frac{\partial\mbh}{y} + \frac{2}{\mbh}\frac{\partial\mlc}{y}. 
\end{eqnarray}
Combining \autoref{eq:appendix1}, \ref{eq:appendix2}, and \ref{eq:appendix3} we find, 
\begin{equation}
    \label{eq:appendix4}
    \frac{\partial\mbh}{\mbh} = \frac{\mtot}{\mtot+2\mlc} \frac{\partial x}{x} + \frac{2\mlc}{\mtot+2\mlc} \frac{\partial\mlc}{\mlc} 
\end{equation}
Using \autoref{eq:appendix4} and \ref{eq:mf-fractional}, replacing $x$, equating differentials with errors, and taking absolute values of possible errors, we find \autoref{eq:delta-mbh}. 
\end{document}